\documentclass[aps,prb,twocolumn,superscriptaddress,longbibliography,floatfix]{revtex4-2}

\usepackage{float,amsmath,amssymb,bbm,mathrsfs,bm,braket,color,graphicx,comment,amsfonts,dsfont,mathrsfs}
\usepackage{xcolor}
\usepackage[colorlinks,citecolor=blue,urlcolor=blue]{hyperref}
\usepackage[mathcal]{euscript}
\usepackage[normalem]{ulem}
\usepackage{comment}
\usepackage{dsfont}
\usepackage{xfrac}
\usepackage{graphicx} 

\DeclareMathAlphabet\mathbfcal{OMS}{cmsy}{b}{n}

\renewcommand{\vec}[1]{\bm{#1}}
\newcommand{\dd}{\mathrm{d}}

\begin{document}
	
\title{Fluctuation induced piezomagnetism in local moment altermagnets}

\author{Kostiantyn V. Yershov}
\affiliation{Leibniz-Institut f\"{u}r Festk\"{o}rper- und Werkstoffforschung, Helmholtzstraße 20, D-01069 Dresden, Germany}
\affiliation{Bogolyubov Institute for Theoretical Physics of the National Academy of Sciences of Ukraine, 03143 Kyiv, Ukraine}

\author{Volodymyr P. Kravchuk}
\affiliation{Leibniz-Institut f\"{u}r Festk\"{o}rper- und Werkstoffforschung, Helmholtzstraße 20, D-01069 Dresden, Germany}
\affiliation{Bogolyubov Institute for Theoretical Physics of the National Academy of Sciences of Ukraine, 03143 Kyiv, Ukraine}

\author{Maria Daghofer}
\affiliation{Institut f\"{u}r Funktionelle Materie und Quantentechnologien, Universit\"{a}t Stuttgart, 70550 Stuttgart, Germany}

\author{Jeroen van den Brink}
\affiliation{Leibniz-Institut f\"{u}r Festk\"{o}rper- und Werkstoffforschung, Helmholtzstraße 20, D-01069 Dresden, Germany}
\affiliation{Institute of Theoretical Physics and W{\"u}rzburg-Dresden  Cluster of Excellence {\it ct.qmat}, Technische Universit{\"a}t Dresden, 01062 Dresden, Germany} 

\date{\today}

\begin{abstract}
It was recently discovered that, depending on their symmetries, collinear antiferromagnets may break spin degeneracy in momentum space, even in absence of spin-orbit coupling. Such systems, dubbed altermagnets, have electronic bands with a spin-momentum texture set mainly by the combined crystal-magnetic symmetry. This discovery motivates the question which novel physical properties derive from altermagnetic order. Here we show that one consequence of altermagnetic order is a fluctuation-driven piezomagnetic response. Using two Heisenberg models of $d$-wave altermagnets, a checkerboard one and one for rutiles,  we determine the fluctuation induced piezomagnetic coefficients considering temperature induced transversal spin fluctuations. We establish in addition that magnetic fluctuations induce an anisotropic thermal spin conductivity.
\end{abstract}
\maketitle

\section{Introduction}

Altermagnetism has recently emerged as a new type of magnetic ordering, distinct from anti-, ferri- and ferromagnetism. Similarly to collinear antiferromagnets~(AFMs), the net magnetization of a collinear altermagnet~(AM) vanishes by symmetry. They differ from AFMs because the enforcing symmetry is not merely an inversion or translation that connects the two magnetic sub-lattices, but also involves a rotation~\cite{Smejkal20,Smejkal22}. The symmetry of a traditional AFMs render their non-relativistic electronic band structure spin degenerate at all momenta, but the rotation between the two magnetic sublattices in AMs breaks this global degeneracy. The energy scale that governs the splitting between the spin up and down bands in AMs is the local exchange field, which is generally much larger than the relativistic spin-orbit coupling energy scale.
 
The promise of spin current generation due to the spin splitting of bands, renders AMs interesting spintronics and a substantial class of AM material candidates have been identified~\cite{Smejkal22a,Naka19,Yuan20,Naka21,Guo23,Sato23}. Few are metallic (e.g. RuO$_2$ and  CrSb), but by far most are robust insulators, in particular strongly correlated ones (e.g. CoF$_2$). These Mott-type insulators have localized moments and their low elementary excitations are charge neutral magnons. 

The recently developed Landau theory of altermagnetism~\cite{McClarty24} allows to relate the formation of antifertomagnetic N{\'e}el  order directly to key observables such as magnetization, anomalous Hall conductivity, magneto-optic and magneto-elastic probes. It establishes in particular the presence of piezomagnetism, in a situation where spin-orbit coupling is absent. In a piezomagnetic system, a net magnetic moment may be induced by applying mechanical stress, or vice versa, a physical deformation by applying a magnetic field. This response is governed by a trilinear coupling between strain, ferromagnetic magnetization and the N{\'e}el order parameter~\cite{McClarty24,Aoyama24}. This standard free energy description implies that fluctuations of the N{\'e}el order parameter appear to diminish the altermagnetic piezomagnetic response.

In contrast to this, we here show that in localized altermagnets transversal spin fluctuations rather have the opposite effect and actually are the drivers of a piezomagnetic response. To be concrete, we use two Heisenberg models of $d$-wave altermagnet and show that while the fluctuation-driven piezomagnetic response is exponentially small at low temperature when the magnetic modes are gapped, it increases with temperature and thus by thermal magnon occupation, reaching a maximum close to the critical temperature. Analytical expressions for this fluctuation induced piezomagnetic response compare well with numerical simulations in which the magnetic system evolves in time according to the stochastic Landau-Lifshitz-Gilbert equations. Since the applied strain breaks the sublattice symmetry for the considered altermagnetic models, our result confirms a recent prediction of the fluctuation-induced ferrimagnetism in sublattice-imbalanced antiferromagnets~\cite{Consoli21}. 

We also show that the presence of magnetic fluctuations induces a thermal spin conductivity, which is due to the different magnon branches carrying opposite magnetic moment, thus coupling the spin carried by the heat current to the direction of that current.

\begin{figure}
	\includegraphics[width=.9\columnwidth]{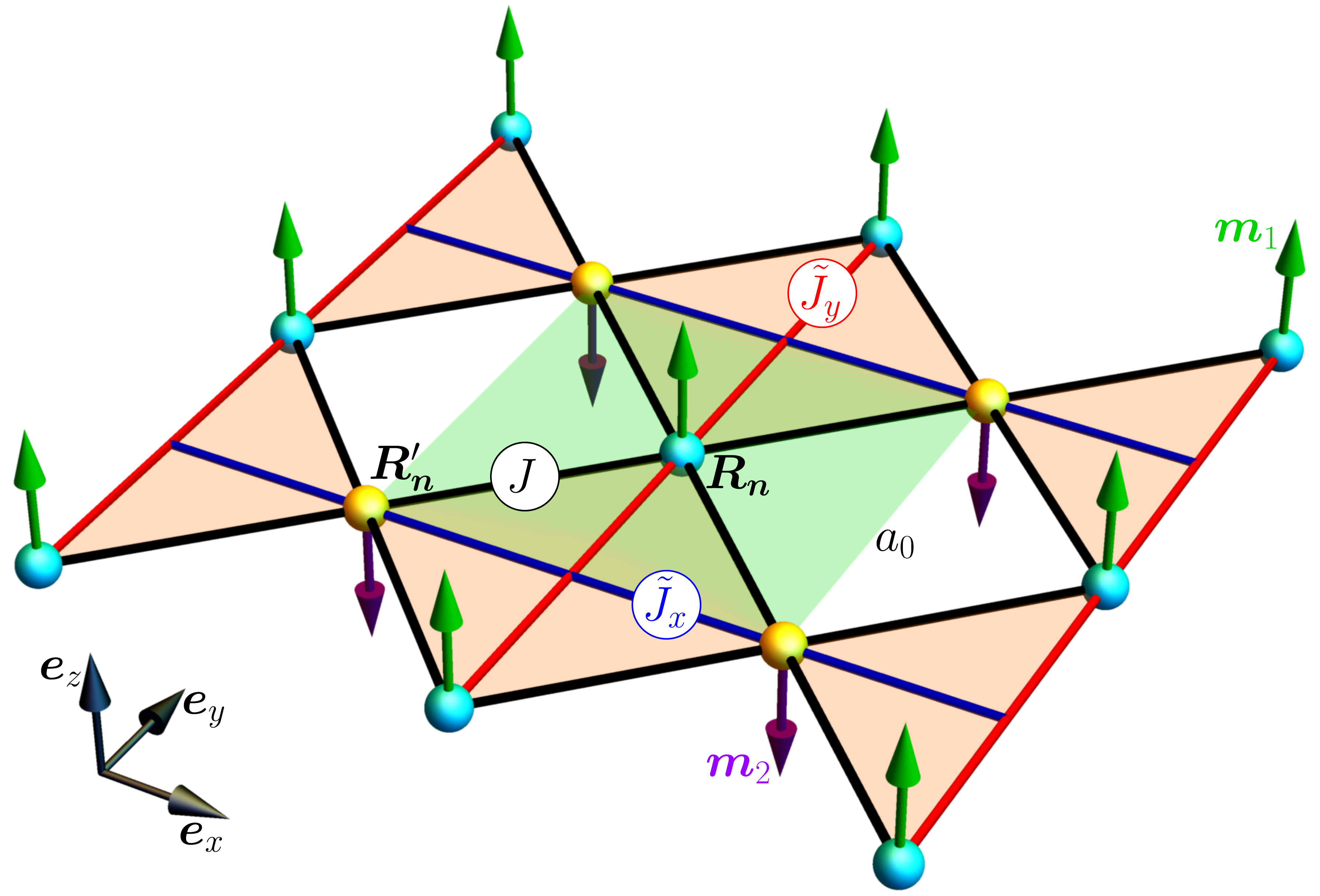}
	\caption{Representation of the Heisenberg checkerboard Hamiltonian in Eq.\eqref{eq:H} which consists of two antiferromagnetically coupled square sublattices of magnetic moments $\mu_s\vec{m}_1(\vec{R}_{\vec{n}})$ and $\mu_s\vec{m}_2(\vec{R}'_{\vec{n}})$ with $\vec{m}_{1,2}$  unit vectors. In addition to the AFM Heisenberg exchange of strength $J$ acting between the nearest neighbors (black bonds), extra Heisenberg interactions of strength $\tilde{J}_x$ (blue bonds) and $\tilde{J}_y$ (red bonds) act along the corresponding diagonals within the sublattices $\vec{m}_2$ and $\vec{m}_1$, respectively. We consider the range of parameters corresponding to the compensated AFM ground state with the antiparallel magnetization of the sublattices. The primitive unit cell of the considered two-sublattice system is shown by the green square with side $a_0$.}\label{fig:checkerboard}
\end{figure}

\section{Heisenberg models of $d$-wave altermagnets}\label{sec:main}
We start from the AM Heisenberg checkerboard Hamiltonian~\cite{Canals02}
as illustrated in Fig.~\ref{fig:checkerboard}. We consider two square sublattices of discrete classical magnetic moments $\vec{\mu}_1(\vec{R}_{\vec{n}})$ and $\vec{\mu}_2(\vec{R}'_{\vec{n}})$ of fixed magnitude located in the positions determined by the Bravais vectors of each of the sublattices, namely $\vec{R}_{\vec{n}}=a_0(n_x\vec{e}_x+n_y\vec{e}_y)$ and $\vec{R}'_{\vec{n}}=\vec{R}_{\vec{n}}-\frac{a_0}{2}(\vec{e}_x+\vec{e}_y)$ with $\vec{n}=\{n_x,n_y\}\in\mathbb{Z}\times\mathbb{Z}$, see Fig.~\ref{fig:checkerboard}. Here $a_0$ denotes size of the square primitive unit cell of the two-sublattice system. We assume that all magnetic moments have equal amplitude $\mu_s$, and therefore, it is instructive to introduce the dimensionless unit vectors $\vec{m}_{1,2}=\vec{\mu}_{1,2}/\mu_s$ showing the magnetic moments orientation.

The dynamics of the system under consideration is governed by the set of coupled Landau-Lifshitz equations $\partial_t\vec{m}_{1,2}=\frac{\gamma}{\mu_s}\left[\vec{m}_{1,2}\times\partial\mathcal{H}/\partial\vec{m}_{1,2}\right]$ where $\gamma>0$ is gyromagnetic ratio, number of equations is equal to the number of magnetic moments, and the coupling is provided by the Hamiltonian
\begin{align}\label{eq:H}
	&\mathcal{H}=J\sum\limits_{\langle\vec{R}_{\vec{n}},\vec{R}'_{\vec{n}}\rangle}\vec{m}_1(\vec{R}_{\vec{n}})\cdot\vec{m}_2(\vec{R}'_{\vec{n}})\\
	\nonumber&+\sum\limits_{\vec{R}_{\vec{n}}}\Bigl[\tilde{J}_y\vec{m}_1(\vec{R}_{\vec{n}})\cdot\vec{m}_1(\vec{R}_{\vec{n}}+a_0\vec{e}_y)-Km_{1z}^2-B\mu_sm_{1z}\Bigr]\\
	\nonumber&+\sum\limits_{\vec{R}'_{\vec{n}}}\Bigl[\tilde{J}_x\vec{m}_2(\vec{R}'_{\vec{n}})\cdot\vec{m}_2(\vec{R}'_{\vec{n}}+a_0\vec{e}_x)-Km_{2z}^2-B\mu_sm_{2z}\Bigr].
\end{align}
Here $J>0$, and, in the first sum, $\vec{R}'_{\vec{n}}$ counts the nearest neighbors of $\vec{R}_{\vec{n}}$. Amplitudes of the diagonal interactions $|\tilde{J}_{x,y}|\ll J$ can generally be different, in this way we take into account the deviation from the altermagnetic limit caused by applied mechanical stress. Additionally, we take into account perpendicular easy-axial anisotropy ($K>0$) and interaction with the magnetic field $\vec{B}=B\vec{e}_z$. Here $m_{1z}=\vec{m}_1(\vec{R}_{\vec{n}})\cdot\vec{e}_z$ and $m_{2z}=\vec{m}_2(\vec{R}'_{\vec{n}})\cdot\vec{e}_z$ with $\vec{e}_z=\vec{e}_x\times\vec{e}_y$.

Assuming that in the ground state magnetic moments are collinear to $\vec{e}_z$, we linearize the Landau-Lifshitz equations with respect to the perpendicular components $m_{1,x}$, $m_{1,y}$, $m_{2,x}$, and $m_{2,y}$ and obtain the following dispersion relation for the linear excitations (magnons)
\begin{subequations}\label{eq:disp}
	\begin{align}
		\label{eq:disp-omega}&\omega_\nu(\vec{k})=\omega_0\left(F_{\vec{k}}\pm\Omega_{\vec{k}}^-\right)\pm\gamma B,\\
	\label{eq:disp-F}	&F_{\vec{k}}=\sqrt{\left(1+\frac{\kappa}{2}-\Omega_{\vec{k}}^+\right)^2-\cos^2\frac{k_xa_0}{2}\cos^2\frac{k_ya_0}{2}},\\ 
 \label{eq:disp-Omega}&\Omega_{\vec{k}}^\pm=\frac{\epsilon_x\sin^2\frac{k_xa_0}{2}\pm\epsilon_y\sin^2\frac{k_ya_0}{2}}{2},
	\end{align}
\end{subequations}
for details see Appendix~\ref{app:disp}. Here index $\nu$ numerates two branches corresponding to the signs `$+$' and `$-$' in the right hand side. Frequency $\omega_0=4J\gamma/\mu_s$ determines the typical time scale of the system. $\kappa=K/J$ is the normalized anisotropy, and $\epsilon_\alpha=\tilde{J}_\alpha/J$ with $\alpha=x,y$. In the particular case $\kappa=0$ and $B=0$, dispersion \eqref{eq:disp} reproduces the previously obtained result~\cite{Canals02}.
\begin{figure}
	\includegraphics[width=\columnwidth]{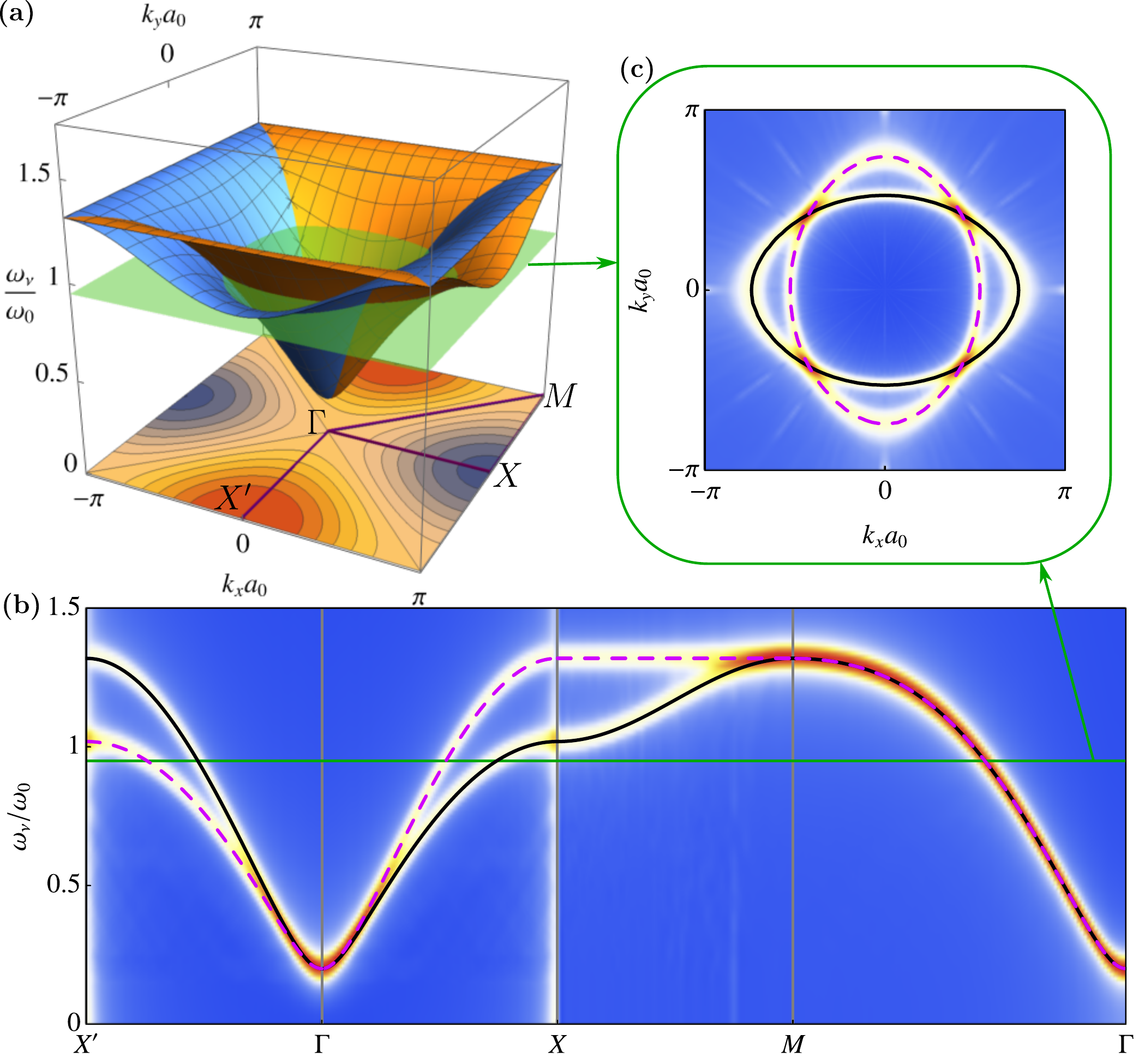}
	\caption{(a) -- Dispersion relation \eqref{eq:disp} within the 1st Brillouin zone for the case $\epsilon_x=\epsilon_y=-0.3$, $\kappa=0.04$, and $B=0$. The value of the splitting between branches is shown by the color code at the bottom. (b) -- the comparison of the dispersion \eqref{eq:disp} with the magnon dispersion obtained using the numerical simulations, for detail see Appendix~\ref{app:simuls}. (c) -- 2D ``surfaces'' of constant energy $\omega_\nu=0.95\omega_0$,}\label{fig:disp}
\end{figure}
An example of the dispersion relation \eqref{eq:disp} is shown in Fig.~\ref{fig:disp}, which demonstrates the anisotropic (in $k$-space) splitting of the magnon branches typical for the d-wave altermagnets~\cite{Smejkal22a}, which for the checkerboard AM are rotated with respect to each other by $\pi/2$. Note the very good agreement with the spectra obtained by means of the numerical simulations, for details see Fig.~\ref{fig:disp}(b) and Appendix~\ref{app:simuls}. Here and below we use negative $\tilde{J}_\alpha$ to avoid frustration in the system. However, independent of the sign of $\tilde{J}_\alpha$, the presented theory is valid under the condition $|\tilde{J}_{\alpha}|\ll J$, which preserves the perpendicular AFM state as the ground state.

\begin{figure}
    \centering
    \includegraphics[width=\columnwidth]{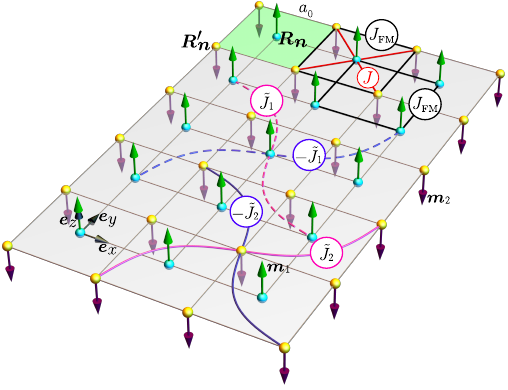}
    \caption{ Model of a rutile bilayer film. Magnetic atoms belonging to two different layers (and different sublattices $\vec{m}_{1,2}$ in the same time) are shown by blue and yellow balls. The nearest neighbours antiferromagnetic ($J$) and ferromagnetic ($J_{\textsc{fm}}$) Heisenberg exchange couplings act between sublattices and within each of the sublattices, respectively. The altermagnetic properties are modeled by the next-nearest exchange couplings $\tilde{J}_{1,2}$ acting within each of the sublattices. The magnetic unit cell is shown by the green square with side $a_0$.}
    \label{fig:rutile}
\end{figure}

\begin{figure}
	\includegraphics[width=\columnwidth]{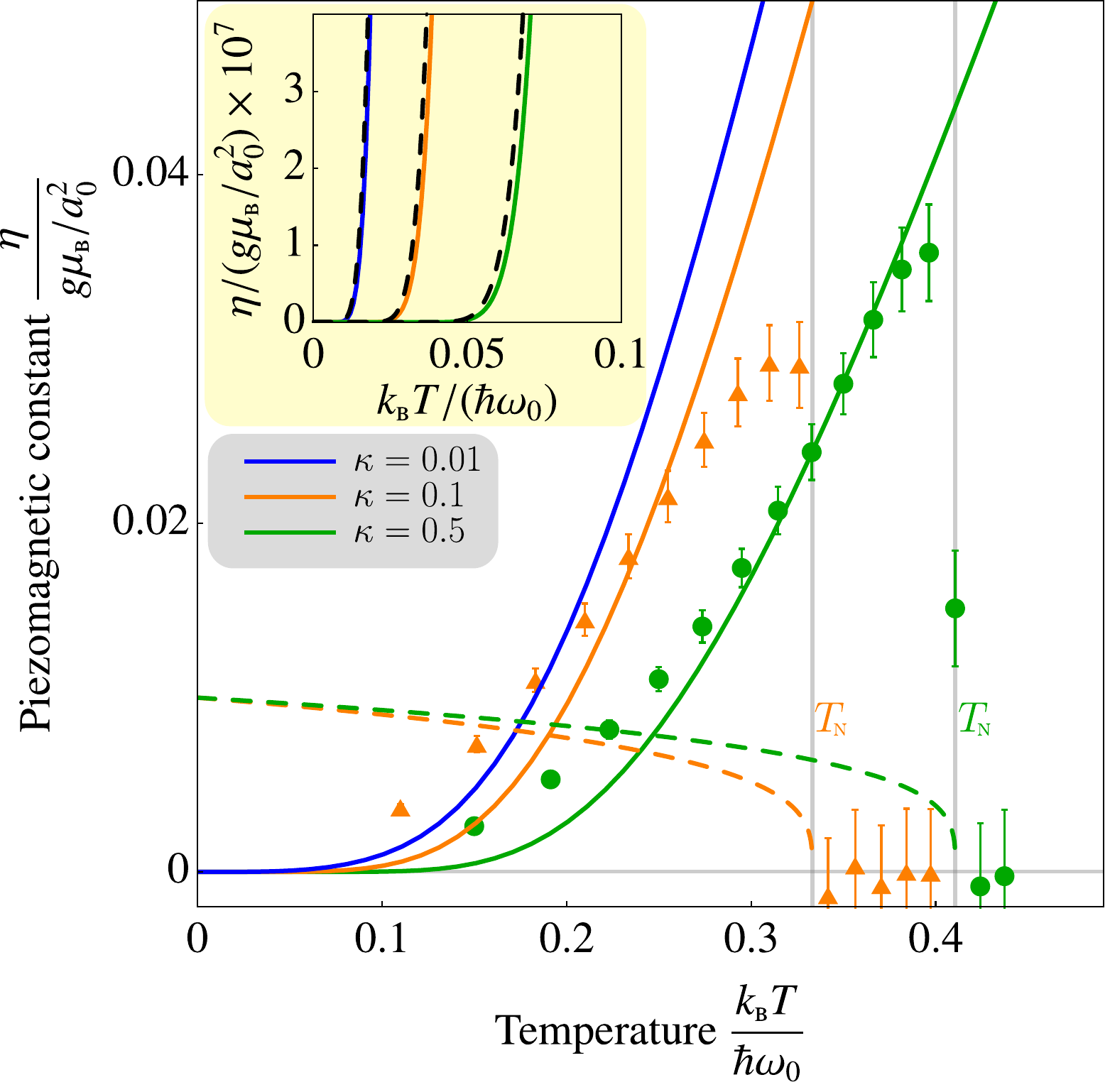}
	\caption{The fluctuation induced piezomagnetic coupling constant $\eta$ as a function of temperature is shown for $\epsilon_x=-0.4$, $\epsilon_y=-0.2$, and for three different anisotropy values.  Symbols correspond to the results of numerical simulations. Dashed lines show the temperature dependence of the normalized ferrimagnetic piezomagnetic constant, $\eta_{\textsc{fm}}/(g\mu_{\textsc{b}}/a_0^2)\propto|\vec{n}|$ for $\mu_s/(g\mu_{\textsc{b}})=0.02$, see Appendix~\ref{app:cont-model}. The temperature dependence of $|\vec{n}|$ is the same as in Fig.~\ref{fig:N-vs-T}. N{\'e}el temperatures corresponding to different anisotropies are indicated by vertical gray lines. The inset demonstrates the asymptotic behavior~\eqref{eq:M-asym} (dashed lines) in the limit of low temperature.}\label{fig:M-vs-T}
\end{figure}

As an alternative to the checkerboard, we consider a model of bilayer rutile film shown in Fig.~\ref{fig:rutile}. Here we generalize the previously proposed~\cite{Gomonay24a} model of rutile assuming that the altermagnetic bonds in different sublattices are generally different $\tilde{J}_1\ne\tilde{J}_2$. In this manner, we model an effect of strain applied one of the lattice diagonals. Details of the rutile Hamiltonian are explained Appendix~\ref{app:rutiles}. 
Applying the same routine as for the checkerboard model Hamiltonian (see Appendix~\ref{app:disp}) we obtain that the dispersion relation of magnons excited over the perpendicular AFM state (shown in Fig.~\ref{fig:rutile}) has the same form as Eq.~\eqref{eq:disp-omega} with
\begin{subequations}\label{eq:F-Omega-rutile}
    \begin{align}
        &F_{\vec{k}}=\biggl\{\biggl[1+\frac{\kappa}{2}+\upsilon\left(\sin^2\frac{k_xa_0}{2}+\sin^2\frac{k_ya_0}{2}\right)\\\nonumber
&+\delta\epsilon\sin(k_xa_0)\sin(k_ya_0)\biggr]^2-\cos^2\frac{k_xa_0}{2}\cos^2\frac{k_ya_0}{2}\biggr\}^{1/2}\\ 
        &\Omega_{\vec{k}}^-=\bar{\epsilon}\sin(k_xa_0)\sin(k_ya_0).
    \end{align}
\end{subequations}
 Here $\upsilon=J_{\textsc{fm}}/J$ and we defined $\delta\epsilon=(\epsilon_1-\epsilon_2)/2$,  $\bar\epsilon=(\epsilon_1+\epsilon_2)/2$ with $\epsilon_\alpha=\tilde{J}_\alpha/J$. The definitions of $\kappa$ and $\omega_0$ are the same as in Eq.~\eqref{eq:disp-omega}.

\subsection{Fluctuation induced piezomagnetism}\label{sec:eta}
We now introduce the magnetic moment as a thermodynamic quantity $M=-\partial_B\mathcal{F}$~\cite{Ashcroft76,Aharoni96}, where $\mathcal{F}$ is the Helmholtz free energy and $B$ is the applied magnetic field~\footnote{$M$ is the magnetic moment along the applied field $B$.}. For low enough temperature, magnons can be considered as gas on noninteracting bosons, in this case~\cite{LandauIX,Akhiezer68} $\mathcal{F}=E_0+\frac{1}{\beta}\sum_{\vec{k},\nu}\ln\left(1-e^{-\beta E_{\vec{k},\nu}}\right)$, where $E_0$ is energy of the ground state, $\beta=1/(k_{\textsc{b}}T)$ is inverse temperature and $E_{\vec{k},\nu}=\hbar\omega_\nu(\vec{k})$ is energy of a magnon. The summation is performed over $\vec{k}$-vectors within the 1st Brillouin zone. Taking into account that the energy $E_0$ of the considered ground state does not depend on the applied field, we differentiate the free energy and present the magnetic moment in form
\begin{equation}\label{eq:M-gen}
	M=\sum\limits_{\vec{k},\nu}\frac{\mu_{\vec{k},\nu}}{e^{\beta E_{\vec{k},\nu}}-1},
\end{equation}
which enables one to recognize the quantity $\mu_{\vec{k},\nu}=-\partial_BE_{\vec{k},\nu}$ as magnetic moment of one magnon~\cite{Ashcroft76}. Note that according to \eqref{eq:disp}, one has $\mu_{\vec{k},\nu}=\mp\gamma\hbar=\mp g\mu_{\textsc{b}}$ with $g$ being the $g$-factor and $\mu_{\textsc{b}}>0$ being the Bohr magneton. Thus, the magnons belonging to different branches carry magnetic moments of opposite signs. With the use of \eqref{eq:disp} and \eqref{eq:M-gen} we derive the following expression for the magnetic moment density in vanishing applied magnetic field
\begin{equation}\label{eq:M-den}
	\mathcal{M}=\frac{g\mu_{\textsc{b}}}{a_0^2}\!\!\iint\limits_{-\pi}^{\;\;\;\;\;\pi}\!\!\frac{\dd q_x\dd q_y}{(2\pi)^2}\frac{\sinh\left(\beta'\Omega_{\vec{q}}^-\right)}{\cosh\left(\beta'F_{\vec{q}}\right)-\cosh\left(\beta'\Omega_{\vec{q}}^-\right)},
\end{equation}
where we introduced the magnetization $\mathcal{M}=M/(L_xL_y)$ with $L_xL_y$ being the altermagnet area, $\vec{q}=a_0\vec{k}$ is the dimensionless wave-vector, $\beta'=\hbar\omega_0/(k_{\textsc{b}}T)$, and we proceed from the summation to integration over the 1st Brillouin zone, assuming the large size of the magnet. Expression \eqref{eq:M-den} is universal for both checkerboard and rutile models, but different definitions of $F_{\vec{k}}$ and $\Omega_{\vec{k}}^-$ provided in \eqref{eq:disp} and \eqref{eq:F-Omega-rutile}, respectively. 

For the checkerboard system, using~\eqref{eq:disp}, one can easily show that $\mathcal{M}\to-\mathcal{M}$ under the interchange $\epsilon_x\leftrightarrow\epsilon_y$. As a consequence, one has $\mathcal{M}=0$ for $\epsilon_x=\epsilon_y$, i.e. the magnetization vanishes in the AM limit. This property motivates us to write $\mathcal{M}=\eta(\epsilon_x-\epsilon_y)$ introducing a coefficient $\eta$ which for convenience we call the piezomagnetic constant \footnote{Strictly speaking, $\eta$ is a part of the piezomagnetic coefficient $\eta_{\text{piezo}}=\eta(\partial\delta\epsilon/\partial\varepsilon)$, where $\delta\epsilon=\epsilon_x-\epsilon_y$ and $\varepsilon$ is the applied strain. Consideration of the quantity $(\partial\delta\epsilon/\partial\varepsilon)$ is beyond the scope of this paper.}. Utilizing this terminology we assume the linear relation between the applied infinitesimal strain and the change of the exchange integrals, implying the basic mechanism of 
the monotonic (exponential) decay of the direct exchange between localized electrons with the distance~\cite{LandauIII}. However, in some materials (e.g. CrI$_3$~\cite{Luo23}) the dependence of some exchange integrals on the inter-atomic distance can be nonmonotonous implying the importance of the nonlinear terms.  

The temperature evolution of $\eta$ is shown in Fig.~\ref{fig:M-vs-T}. For the limit $T\to0$, it is exponentially suppressed, which is a typical behavior for a gapped system. Anisotropy strengthens the magnetization stiffness of the sublattices, suppressing thermal occupation of magnons and thus the emergent magnetic moment. This is more clearly shown in the inset of Fig.~\ref{fig:M-vs-T}. In the limit of low temperatures ($\beta'/\Delta\gg1$), we estimate integral \eqref{eq:M-den} by means of the Laplace method~\cite{Bender99}:
\begin{equation}\label{eq:M-asym}
	\eta\approx\frac{g\mu_{\textsc{b}}}{a_0^2}\frac{2}{\pi}f(\epsilon_x,\epsilon_y)\frac{\Delta^2}{\beta'}\ln\frac{1}{1-e^{-\beta'\Delta}},
\end{equation}
where $\Delta=\sqrt{\kappa(1+\frac{\kappa}{4})}$ is the gap size in units $\hbar\omega_0$. Function $f$ is defined in \eqref{eq:eta-f}. In the limit of small altermagnetism ($\epsilon_\alpha\ll1$), we approximate $f\approx1+O(\epsilon_x+\epsilon_y)$. For details see Appendix~\ref{app:approx}.

\label{eta-rutile}For the rutile model, the temperature dependence of the piezomagnetic coefficient $\eta=\mathcal{M}/(2\delta\epsilon)$ is qualitatively similar to the dependencies shown in Fig.~\ref{fig:M-vs-T}. The low-temperature asymptotic coincides with \eqref{eq:M-asym} up to the value of function $f=O(\epsilon_1+\epsilon_2)$, for details see Appendix~\ref{app:rutiles}. Thus, in the limit of small altermagnetism, one has $\mathcal{M}\propto\tilde{J}_1^2-\tilde{J}_2^2$ and $\mathcal{M}\propto\tilde{J}_x-\tilde{J}_y$ for the rutile and checkerboard models, respectively. And this is the only principle difference between piezomagnetic properties of these models. 

\label{RuO2-estimation}As an example of the altermagnet with the symmetry of rutile, we consider  RuO$_2$ with the parameters listed in supplemental materials of Ref.~\onlinecite{Gomonay24a} and estimate $\eta\approx0.13\mu_{\textsc{b}}/\mathrm{nm}^2$ for the room temperature.

To verify our analytical prediction, we perform spin-lattice simulations for the checkerboard model. The simulations are based on the classical Landau-Lifshitz equations with the thermal fluctuations included. Details of the numerical routine are explained in Appendix~\ref{app:simuls}. One should note a nice agreement with analytics in the intermediate range of temperatures, see Fig~\ref{fig:M-vs-T}. However, in vicinity of the N{\'e}el temperature $T\lessapprox T_{\textsc{n}}$, the simulated piezomagnetic constant drops down reflecting the vanishing of the order parameter $|\vec{n}|$, see Fig.~\ref{fig:N-vs-T}. This is in contrast to the analytical result \eqref{eq:M-den} which was derived for the ground state with $|\vec{n}|=1$. Furthermore, we treat magnons as a gas of non-interacting bosons, which is not true at high temperatures. Since the simulations are based on the purely classical dynamics of magnetic moments, the Bose-Einstein statistic for magnons is not incorporated. This gives rise to small discrepancies for small temperatures.

Using the Landau phenomenological theory of altermagnetism~\cite{McClarty24}, one can show that the situation $\tilde{J}_x\ne\tilde{J}_y$ leads to ferrimagnetism, i.e. $|\vec{m}_1|\ne|\vec{m}_2|$, with the corresponding piezomagnetic constant $\eta_{\textsc{fm}}=\mu_s|\vec{\mathrm{n}}|/(2a_0^2)$, see Appendix~\ref{app:cont-model}. Using the spin-lattice simulations, we determine the temperature evolution of $|\vec{n}|$ and fit it by the dependence $|\vec{n}|_{\mathrm{fit}}=(1-T/T_{\textsc{n}})^\alpha$, see Fig.~\ref{fig:N-vs-T}. The temperature dependencies of $\eta_{\textsc{fm}}$ is shown in Fig.~\ref{fig:M-vs-T} by dashed lines. One can see that for the small enough ratio $\mu_s/(g\mu_{\textsc{b}})$, the fluctuation-induced piezomagnetism dominates in a wide region of high temperatures.

The gapless limit ($\kappa=0$) is not achievable in the 2D geometry because of the Mermin-Wagner theorem. To consider this limit, we performed a simple 3D generalization of the checkerboard model in which  we consider a stacked system composed of the 2D checkerboard layers shown in Fig.~\ref{fig:checkerboard} with ferromagnetic inter-layer exchange coupling $J_z$, see Appendix~\ref{app:3D}. The corresponding low-temperature asymptotic for $\eta^{\text{3D}}$ is presented in Eqs.~\eqref{eq:eta-3D}. For the gapless case we estimate $\eta^{\text{3D}}\propto T^4$, namely
\begin{equation}\label{eq:eta-3D-gpls}
    \eta^{\text{3D}}\approx\frac{g\mu_{\textsc{b}}}{a_0^2c_0}\frac{8\sqrt{2}\pi^{2}}{45\sqrt{\upsilon_z}}\frac{1}{\beta'^4}.
\end{equation}
Here $\upsilon_z=J_z/J$, $c_0$ is the distance between the stacked layers, and we assumed that $k_{\textsc{b}}T\ll\sqrt{\upsilon_z}\hbar\omega_0$. Note that the $T^4$-dependence of the fluctuation-induced magnetization is typical for collinear Heisenberg antiferromagnets with the broken sublattice symmetry~\cite{Consoli21}.

\subsection{Fluctuation induced thermal spin conductivity}\label{sec:sigma}
Let us now consider the possibility of the generation of spin current $j$ by the applied temperature gradient $\vec{\nabla}T$:
\begin{equation}
    j_\alpha=\sigma_{\alpha\beta}\partial_\beta T,
\end{equation}
where $\sigma_{\alpha\beta}$ is tensor of thermal spin conductivity. Within the relaxation time approximation \cite{Kittel05} we obtain
\begin{equation}\label{eq:sigma}
\sigma_{\alpha\beta}=\frac{\tau_{\text{rlx}}}{L_xL_y}\sum\limits_{\vec{k},\nu}c_{\vec{k},\nu}\left(v_{\vec{k},\nu}\right)_\alpha\left(v_{\vec{k},\nu}\right)_\beta,
\end{equation}
where $\tau_{\text{rlx}}$ is the relaxation time -- the average time between magnons collisions, $\left(v_{\vec{k},\nu}\right)_\alpha=\partial\omega_{\nu}(\vec{k})/\partial k_\alpha$ is group velocity, and
\begin{equation}\label{eq:cap}
	c_{\vec{k},\nu}=\partial_T\frac{\mu_{\vec{k},\nu}}{e^{\beta E_{\vec{k},\nu}}-1}
\end{equation}
is spin capacity per magnon. 
For the dispersion relation \eqref{eq:disp-omega}, the general form of the thermal spin conductivity tensor defined in \eqref{eq:sigma} is
\begin{equation}\label{eq:sigma-gen}
    \begin{split}
        &\sigma_{\alpha\beta}=-\frac{\gamma k_{\textsc{b}}}{4}\frac{\tau_{\text{rlx}}\omega_0}{L_xL_y}\beta'^2\sum\limits_{\vec{k}\in1.\text{BZ}}\sum\limits_{\nu=\pm1}\\
        &\nu\frac{(F_{\vec{k}}+\nu\Omega_{\vec{k}}^-)(\partial_{k_\alpha}F_{\vec{k}}+\nu\partial_{k_\alpha}\Omega_{\vec{k}}^-)(\partial_{k_\beta}F_{\vec{k}}+\nu\partial_{k_\beta}\Omega_{\vec{k}}^-)}{\sinh^2[\frac{\beta'}{2}(F_{\vec{k}}+\nu\Omega_{\vec{k}}^-)]}.
    \end{split}
\end{equation}
Expression \eqref{eq:sigma-gen} is universal for both checkerboard and rutile models but definitions of $F_{\vec{k}}$ and $\Omega_{\vec{k}}^-$.

\begin{figure}
    \centering
    \includegraphics[width=0.95\columnwidth]{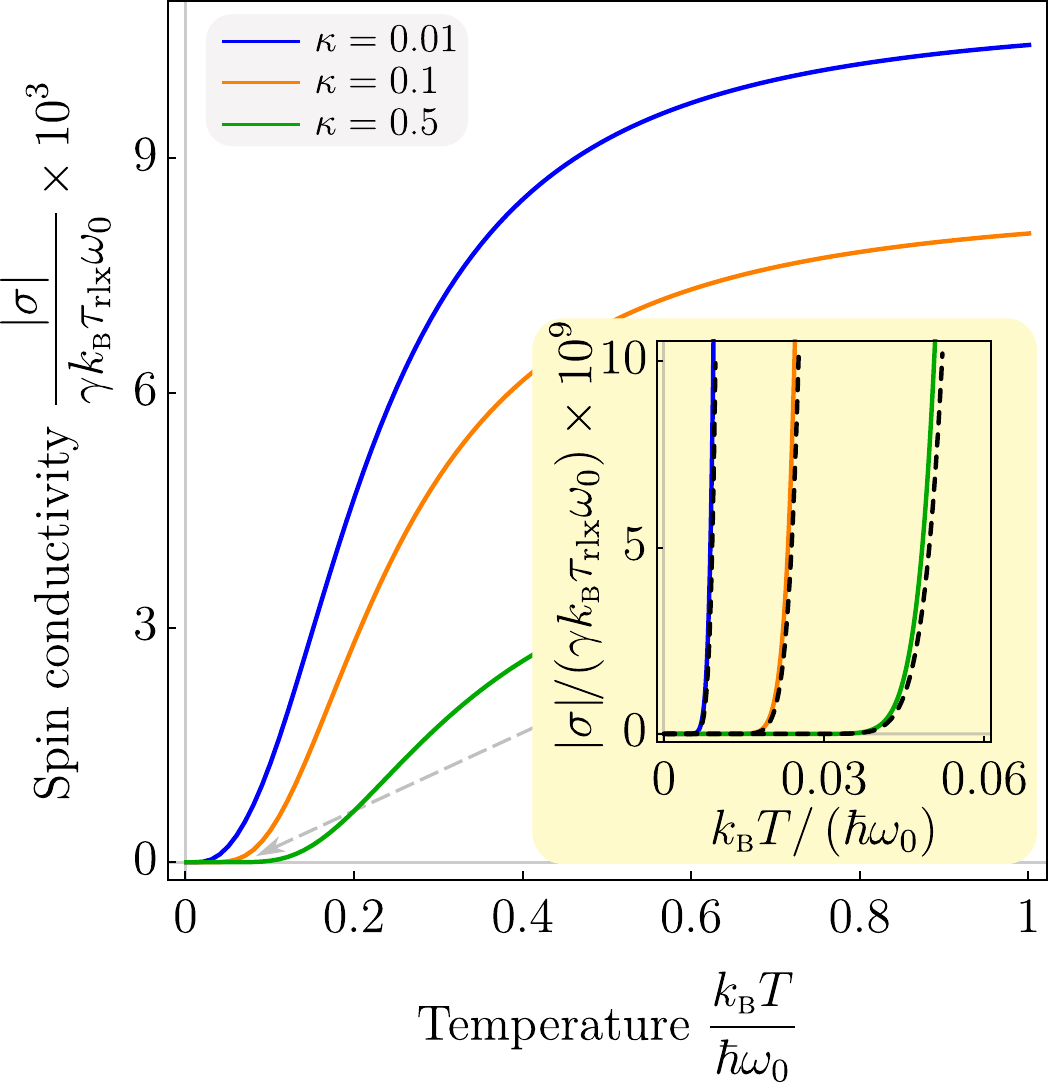}
    \caption{Diagonal elements of the spin conductivity tensor~\eqref{eq:sigma} computed for $\epsilon_x=\epsilon_y=-0.1$ and for three different anisotropy values. The inset demonstrates the asymptotic behavior~\eqref{eq:sigma-ass}~(dashed lines) in the limit of low temperature.}
    \label{fig:sigma}
\end{figure}

For the checkerboard model, $F_{\vec{k}}$, $\Omega_{\vec{k}}^-$ are even and $\partial_{k_\alpha}F_{\vec{k}}$, $\partial_{k_\alpha}\Omega_{\vec{k}}^-$ are odd functions of $k_\alpha$. From \eqref{eq:sigma-gen} we conclude that the diagonal components $\sigma_{xy}=\sigma_{yx}=0$ are averaging out during summation over $k_\alpha$. For the case $\epsilon_x=\epsilon_y$, additional symmetries $F_{\vec{k}}\to F_{\vec{k}}$ and $\Omega^-_{\vec{k}}\to-\Omega^-_{\vec{k}}$ take place under the interchange of variables $k_x\leftrightarrow k_y$. In this case, Eq.~\eqref{eq:sigma-gen} results in $\sigma_{xx}=-\sigma_{yy}$  and the conductivity tensor obtains the following form
\begin{equation}\label{eq:sigma-tensor}
    [\sigma_{\alpha\beta}]=\sigma\begin{bmatrix}
        -1 & 0 \\
        0  & 1
    \end{bmatrix}.
\end{equation}
The temperature dependence of the conductivity amplitude $\sigma$ is shown in Fig.~\ref{fig:sigma}. According to \eqref{eq:sigma-tensor}, spin current $\vec{j}$ flows at angle $\pi-\varphi$ to the direction $\vec{e}_x$ where $\varphi=\angle(\vec{\nabla}T,\vec{e}_x)$.
For the case of low temperature ($\beta'/\Delta\gg1$) we obtain $-\sigma_{xx}\approx\sigma_{yy}\approx\sigma$ with
\begin{equation}\label{eq:sigma-ass}
    \sigma\approx\gamma k_{\textsc{b}}\frac{\omega_0\tau_{\text{rlx}}}{\pi}\Delta^2\bar{\epsilon}\ln\frac{1}{1-e^{-\beta'\Delta}},
\end{equation}
where $\bar{\epsilon}=(\epsilon_x+\epsilon_y)/2$. For the exact expressions for the components of $\sigma_{\alpha\beta}$, see Appendix~\ref{app:sigma}. For a 3D generalization of the checkerboard model we consider the gapless limit in which $\sigma\propto T^3$, namely
\begin{equation}\label{eq:sigma-3D}
    \sigma^{\mathrm{3D}}\approx\gamma k_{\textsc{b}}\frac{\tau_{\text{rlx}}\omega_0\epsilon}{c_0\sqrt{\upsilon_z}\beta'^3}\frac{8\sqrt{2}\pi^2}{75}.
\end{equation}
see Appendix~\ref{app:3D} for details. Here we assumed that $\epsilon_x=\epsilon_y=\epsilon$ and $k_{\textsc{b}}T\ll\sqrt{\upsilon_z}\hbar\omega_0$.

\label{sigma-rutile}For the rutile model, tensor $\sigma_{\alpha\beta}$ has the same structure as \eqref{eq:sigma-tensor} up the rotation of the reference frame by $\pi/4$, i.e. $\sigma_{xx}=\sigma_{yy}=0$ and $\sigma_{xy}=\sigma_{yx}=\sigma$. The low-temperature behavior of $\sigma$ coincides with \eqref{eq:sigma-ass} up to a numerical constant, see Eq.~\eqref{eq:sigma-ass-rtl}. Since the low-temperature asymptotics of $\eta$ and $\sigma$ obtained for both models coincides up to a constant multiplier, we conjecture that the results \eqref{eq:M-asym} and \eqref{eq:sigma-ass} are universal for two-dimansional $d$-wave altermagnets.

\section{Conclusions}
We analyzed the contribution of magnons to the thermodynamic properties an altermagnetic film whose magnetic subsystem is approximated by the checkerboard model. This AM has two important features: (i) it results in the anisotropic (in $k$-space) splitting of the magnon spectra typical for $d$-wave altermagnets, and (ii) it allows an easy relation between the magnetic properties and the applied strain~$\propto(\tilde{J}_x-\tilde{J}_y)$. The Landau theory for altermagnets implies a trilinear coupling between strain, ferromagnetic magnetization and the N{\'e}el order parameter. Therefore an applied strain leads in general to a ferrimagnetic state: a strain that breaks the AM symmetry allows the magnitude of the moments on the two sublattices to become different (see Appendix~\ref{app:cont-model} and Ref.~\cite{McClarty24}). Due to the trilinear coupling this longitudinal response is present in the ground state and vanishes together with the N\'eel parameter when temperature increases. 

Here we have identified a piezomagnetic response that instead grows with temperature, because it is driven by thermally excited magnons and is described by formula~\eqref{eq:M-den}. This piezomagnetic response is due to transversal magnetic fluctuations and can thus also be expected for systems with fixed local moments. It reaches a maximum in a temperature region just below $T_{\textsc{n}}$. Interestingly, the thermally-induced piezomagnetism is dominant for materials with small magnetic moments $\mu_s\ll\mu_{\textsc{b}}$ in the temperature regime $T\lesssim T_{\textsc{n}}$. We have also shown that in presence of magnetic fluctuations a spin current is generated by an applied temperature gradient due to different magnon branches carrying opposite magnetic moment. The spin carried by the heat current couples to the direction of that current. These results are easily generalized to higher dimensions and different lattice geometries. For instance, we show that the model of an altermagnet with a structure of rutile results in the same low-temperature behavior of $\eta$ and $\sigma$ as the checkerboard model.

This fluctuation induced piezomagnetic effect may be of interest for the control of AM domains, which is key for development of AM-based spintronics because macroscopic altermagnetic properties and responses vanish when domains are averaged over. As AM domains are related by time-reversal symmetry~\cite{Aoyama24,Gomonay24a}, their piezomagnetic response has an opposite sign. Thus an energy difference between domains can be induced by applying simultaneously strain and magnetic field in the appropriate directions. Particularly, applying these two during cooling across $T_{\textsc{n}}$ opens an efficient route to favor only one of the domains. As we have shown here precisely in this temperature regime magnetic fluctuations strongly affect the piezomagnetic response.

A way to observe the fluctuation-induced piezomagnetism may be to apply the experimental approach of Ref.~\onlinecite{Aoyama24} to a $d$-wave altermagnet, as for the originally considered $g$-wave altermagnet, the effect is averaged out because of the higher symmetry of the magnon dispersion.

\section*{Acknowledgments}
We thank Jorge Facio and Oleg Jansson for fruitful discussions and Ulrike Nitzsche for technical support. This work was supported by the Deutsche Forschungsgemeinschaft (DFG, German Research Foundation) through the Sonderforschungsbereich SFB 1143, grant No. YE 232/2-1, and under Germany's Excellence Strategy through the W\"{u}rzburg-Dresden Cluster of Excellence on Complexity and Topology in Quantum Matter -- \emph{ct.qmat} (EXC 2147, project-ids 390858490 and 392019).

\appendix
\section{Dispersion relation for the Heisenberg checkerboard model}\label{app:disp}
We start from the linearization of the Landau-Lifshitz equations $\partial_t\vec{m}_{1,2}=\frac{\gamma}{\mu_s}\left[\vec{m}_{1,2}\times\partial\mathcal{H}/\partial\vec{m}_{1,2}\right]$ with respect to the perpendicular components $m_{1,2;x,y}$ on the top of the ground state $\vec{m}_1^0=\vec{e}_z$, $\vec{m}_2^0=-\vec{e}_z$:
\begin{equation}\label{eq:LL-lin}
	\begin{split}
		&\partial_tm_{1\alpha}(\vec{R}_{\vec{n}})=-\varepsilon_{\alpha\beta}\frac{\gamma}{\mu_s}\frac{\partial\mathcal{H}^{(2)}}{\partial m_{1\beta}(\vec{R}_{\vec{n}})},\\
		&\partial_tm_{2\alpha}(\vec{R}'_{\vec{n}})=\varepsilon_{\alpha\beta}\frac{\gamma}{\mu_s}\frac{\partial\mathcal{H}^{(2)}}{\partial m_{2\beta}(\vec{R}'_{\vec{n}})},
	\end{split}
\end{equation}
where $\alpha=x,y$, and the harmonic part of Hamiltonian~\eqref{eq:H} is as follows
\begin{widetext}
\begin{align}
	\nonumber\mathcal{H}^{(2)}=\sum\limits_{\alpha=x,y}\biggl\{&J\!\!\!\sum\limits_{\langle\vec{R}_{\vec{n}},\vec{R}'_{\vec{n}}\rangle}m_{1\alpha}(\vec{R}_{\vec{n}})m_{2\alpha}(\vec{R}'_{\vec{n}})+\sum\limits_{\vec{R}_{\vec{n}}}\Bigl[\tilde{J}_ym_{1\alpha}(\vec{R}_{\vec{n}})m_{1\alpha}(\vec{R}_{\vec{n}}+a_0\vec{e}_y)+\left(2J-\tilde{J}_y+K+\frac{\mu_sB}{2}\right)m_{1\alpha}^2(\vec{R}_{\vec{n}})\Bigr]\\
\label{eq:H2}	&+\sum\limits_{\vec{R}'_{\vec{n}}}\Bigl[\tilde{J}_xm_{2\alpha}(\vec{R}'_{\vec{n}})m_{2\alpha}(\vec{R}'_{\vec{n}}+a_0\vec{e}_x)+\left(2J-\tilde{J}_x+K-\frac{\mu_sB}{2}\right)m_{2\alpha}^2(\vec{R}'_{\vec{n}})\Bigr]\biggr\}
\end{align}
\end{widetext}

Next, we utilize the Fourier transforms on the periodic lattice 
\begin{equation}\label{eq:FT-discr}
	\begin{split}
	f(\vec{R}_{\vec{n}})=\frac{1}{\sqrt{N}}\sum\limits_{\vec{k}\in1.\text{BZ}}\hat{f}(\vec{k})e^{i\vec{k}\cdot\vec{R}_{\vec{n}}},\\
	\hat{f}(\vec{k})=\frac{1}{\sqrt{N}}\sum\limits_{\vec{R}_{\vec{n}}}f(\vec{R}_{\vec{n}})e^{-i\vec{k}\cdot\vec{R}_{\vec{n}}}	
	\end{split}
\end{equation}
supplemented with the completeness relation $\sum_{\vec{R}_{\vec{n}}}e^{i(\vec{k}-\vec{k}')\cdot \vec{R}_{\vec{n}}}=N\delta_{\vec{k},\vec{k}'}$. Here $N$ is the number of magnetic moments in one sublattice. Applying \eqref{eq:FT-discr} to \eqref{eq:LL-lin}, we obtain the equations of motion in reciprocal space
\begin{equation}\label{eq:LL-lin-rec}
	\begin{split}
	&\partial_t\hat{m}_{1\alpha}(\vec{k})=-\varepsilon_{\alpha\beta}\frac{\gamma}{\mu_s}\frac{\partial\mathcal{H}^{(2)}}{\partial\hat{m}_{1\beta}(-\vec{k})},\\
	&\partial_t\hat{m}_{2\alpha}(\vec{k})=\varepsilon_{\alpha\beta}\frac{\gamma}{\mu_s}\frac{\partial\mathcal{H}^{(2)}}{\partial\hat{m}_{2\beta}(-\vec{k})},
\end{split}
\end{equation}
Note that the Fourier transform for the sublattice $\vec{R}'_{\vec{n}}$ coincides with \eqref{eq:FT-discr} up to the replacement $\vec{R}_{\vec{n}}\to \vec{R}'_{\vec{n}}$. In reciprocal space, the harmonic part of he Hamiltonian \eqref{eq:H2} is as follows

\begin{equation}\label{eq:Hk}
	\begin{split}
	&\mathcal{H}^{(2)}=4J\sum\limits_{\vec{k}\in1.\text{BZ}}\biggl[\mathcal{A}_{\vec{k}}\hat{m}_{1\alpha}(\vec{k})\hat{m}_{2\alpha}(-\vec{k})\\
	&+\frac{\mathcal{B}_{\vec{k}}+\mathcal{C}_{\vec{k}}}{2}\hat{m}_{1\alpha}(\vec{k})\hat{m}_{1\alpha}(-\vec{k})+\frac{\mathcal{B}_{\vec{k}}-\mathcal{C}_{\vec{k}}}{2}\hat{m}_{2\alpha}(\vec{k})\hat{m}_{2\alpha}(-\vec{k})\biggr],
	\end{split}
\end{equation}
where the summation over the repeating index $\alpha\in\{x,y\}$ is assumed, $\mathcal{A}_{\vec{k}}=\cos\frac{k_xa_0}{2}\cos\frac{k_ya_0}{2}$, $\mathcal{B}_{\vec{k}}=1+\frac{\kappa}{2}-\Omega_{\vec{k}}^+$, $\mathcal{C}_{\vec{k}}=\Omega_{\vec{k}}^-+b$ with $b=\mu_sB/(4J)$ 
Parameters $\kappa$, $\epsilon_{\alpha}$ as well as $\Omega_{\vec{k}}^\pm$ are defined in the main text.
With \eqref{eq:Hk}, we write Eqs.~\eqref{eq:LL-lin-rec} in the form
\begin{equation}\label{eq:LL-lin-rec1}
	\partial_t\vec{\xi}=\omega_0\mathbb{M}\vec{\xi},
\end{equation}
where $\vec{\xi}=(\hat{m}_{1x},\hat{m}_{1y},\hat{m}_{2x},\hat{m}_{2y})^{\textsc{t}}$, and matrix $\mathbb{M}$ is as follows
\begin{equation}\label{eq:mtrxM}
	\mathbb{M}=\begin{bmatrix}
		0 & -(\mathcal{B}_{\vec{k}}+\mathcal{C}_{\vec{k}}) & 0 & -\mathcal{A}_{\vec{k}} \\
		\mathcal{B}_{\vec{k}}+\mathcal{C}_{\vec{k}} & 0 & \mathcal{A}_{\vec{k}} & 0\\
		0 & \mathcal{A}_{\vec{k}} & 0 & \mathcal{B}_{\vec{k}}-\mathcal{C}_{\vec{k}} \\
		-\mathcal{A}_{\vec{k}} & 0 & -(\mathcal{B}_{\vec{k}}-\mathcal{C}_{\vec{k}}) & 0
	\end{bmatrix}
\end{equation}
 System \eqref{eq:LL-lin-rec1} has solution $\vec{\xi}=\vec{\xi}_0e^{-i\omega t}$, which is nontrivial ($\vec{\xi}_0\ne\vec{0}$) if $\omega=i\omega_0\lambda_\nu$ with $\lambda_\nu$ being the eigenvalues of matrix $\mathbb{M}$. The eigenvalues $\lambda_\nu$ are imaginary and compose two complex-conjugated pairs. The pair of the non-negative eigenfrequencies $\omega_{1,2}=\omega_0(\sqrt{\mathcal{B}_{\vec{k}}^2-\mathcal{A}_{\vec{k}}^2}\pm\mathcal{C}_{\vec{k}})$ coincides with \eqref{eq:disp}. Note that $\gamma B=\omega_0b$.

\section{Numerical simulations of the Heisenberg checkerboard}\label{app:simuls}

\begin{figure*}[t]
	\includegraphics[width=\textwidth]{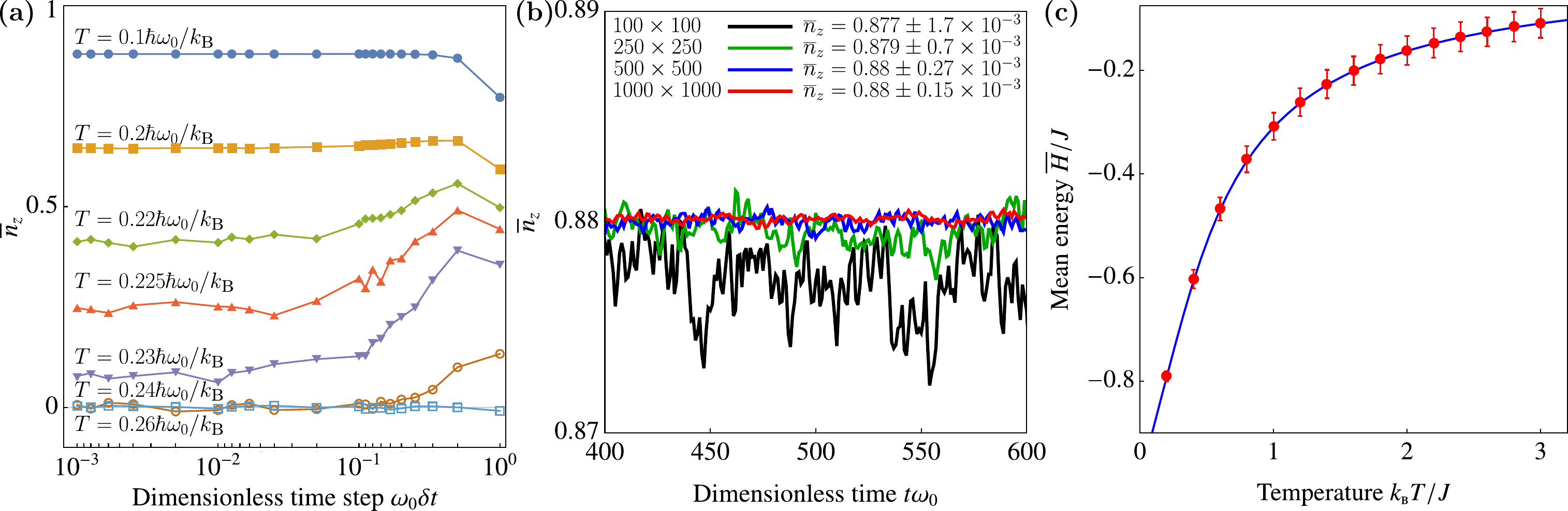}
	\caption{(a) -- Time step dependence of the perpendicular N{\'e}el vector component for different reduced temperatures for Heun's integration scheme. (b) -- Comparison between the dynamics of the perpendicular N{\'e}el vector component $\overline{n}_z(t)$ with different system size. (c) -- Mean energy per spin of the 1D Heisenberg chain, as a function of temperature. Symbols show the result of numerical simulations and the solid line corresponds to the analytical prediction. In all cases simulations were performed for the Gilbert damping parameter $\alpha = 0.1$.}\label{fig:conv}
\end{figure*}

We consider a square lattice with lattice constant $a_0$. Each node is characterized by a magnetic moment $\vec{m}_i$, and index $i$ defines the position of magnetic moment on the lattice with size $N_1\times N_2$. The dynamics of the magnetic system is governed by the stochastic Landau--Lifshitz equations
\begin{equation}\label{eq:LL-sim}
    \begin{split}
    \frac{\mathrm{d}\vec{m}_i}{\mathrm{d}t}&=-\frac{\gamma}{1+\alpha^2}\left[1+\alpha\,\vec{m}_i\times\right]\vec{m}_i\times\vec{H}_i^\textsc{eff},\\
    \vec{H}_i^\textsc{eff}&= -\frac{1}{\mu_s}\frac{\partial\mathcal{H}}{\partial\vec{m}_i} + \vec{H}_i^\textsc{th},
    \end{split}
\end{equation}
where $\alpha$ is a Gilbert damping parameter, $\mathcal{H}$ is defined in~\eqref{eq:H}, and $\vec{H}_i^\textsc{th}$ is a stochastic thermal field given by
\begin{equation}\label{eq:th_field}
    \vec{H}_i^\textsc{th}\left(t\right) = \sqrt{2\mathcal{D}}\vec{\zeta}_i\left(t\right)=\sqrt{2\frac{\alpha\, k_\textsc{b}T}{\gamma\mu_s}}\vec{\zeta}_i\left(t\right),
\end{equation}
where the magnitude is given by the fluctuation-dissipation theorem and $\vec{\zeta}_i$ is white noise, such that the ensemble average and variance of the thermal field fulfill $\left<\vec{H}_{i\alpha}^\textsc{th}(t)\right>=0$ and $\left<\vec{H}_{i\alpha}^\textsc{th}(0)\vec{H}_{j\beta}^\textsc{th}(t)\right>= 2\mathcal{D}\delta_{ij}\delta_{\alpha\beta}\delta(t)$, respectively. To achieve these properties in an implementation, the vectors $\vec{\zeta}_i$ can each be created from three independent standard normally distributed random values at every time step. Note also that in time-integration schemes, to fulfill the fluctuation-dissipation relation, the thermal field needs to be normalized by the time step with a factor $1/\sqrt{\delta t}$.

To evolve a magnetic system in time according to equation~\eqref{eq:LL-sim} we used Heun’s method for temperature-induced effects~\cite{Nowak01,Mentink10,Evans14}, and a fourth-order Runge-Kutta method in other cases. During the integration process, the condition $\left|\vec{m}_i(t)\right| = 1$ is controlled.

\subsection{Convergence of numerical results}\label{sec:sim-CON}
\textit{Finite time step.} For micromagnetic simulations at zero temperature, the minimum time step is a well defined quantity since the largest field (usually the exchange field) essentially defines the precession frequency. However, for spin-lattice simulations using the stochastic Landau--Lifshitz equation, the effective field becomes temperature dependent, see Eqs.~\eqref{eq:LL-sim} and~\eqref{eq:th_field}. The consequence of this is that for spin-lattice models the most difficult region to integrate is near the N{\'e}el temperature. Errors in the integration of the system will be apparent from a non-converged value for the average N{\'e}el vector. This gives a relatively simple case which can then be used to test the stability of integration schemes for the stochastic Landau--Lifshitz equations. A plot of the mean perpendicular component of N{\'e}el vector $n_z$ as a function of temperature is shown in Fig.~\ref{fig:conv}(a) for a system consisting of $500\times 500$ magnetic moments, where we take into account only nearest-neighbors AFM exchange, i.e. $\tilde{J}_x=\tilde{J}_y=0$, and easy-axial anisotropy $\kappa = 0.5$.

The data in Fig.~\ref{fig:conv}(a) shows that for low temperatures reasonably large time steps of $\omega_0\delta t = 10^{-1}$ \footnote{For material parameters of RuO$_2$~\cite{Gomonay24a} the corresponding characteristic time scale parameter is $1/\omega_0\approx 45$ fs.} give the correct solution of the Landau--Lifshitz equations, while near the N{\'e}el temperature the deviations from the correct equilibrium value are significant. Consequently for higher temperatures time steps of $\omega_0\delta t =10^{-3} - 10^{-2}$ are necessary.

\textit{Finite size.} In the next step we check the system size dependence of our results for samples with nearest-neighbors AFM exchange, i.e. $\tilde{J}_x=\tilde{J}_y=0$, and easy-axial anisotropy $\kappa = 0.5$. A plot of the mean perpendicular component of N{\'e}el vector $n_z$ as a function of sample size is shown in Fig.~\ref{fig:conv}(b) for temperature $T = 0.1 \hbar\omega_0/k_\textsc{b}$. The data in Fig.~\ref{fig:conv}(b) shows that a small sample ($100\times100$) gives almost the same value of $\overline{n}_z$ as a large sample ($1000\times1000$). In the following, we consider samples with a size of $500\times500$ magnetic moments.

\textit{1D Heisenberg chain.} The simplest model of classical interacting spins is the 1D Heisenberg chain with nearest-neighbor interactions. For this system, an analytical expression for the mean energy per spin is known~\cite{Fisher64} $\overline{H}/J = k_\textsc{b}T /J - 1/\tanh\left[J/\left(k_\textsc{b}T\right)\right]$. The comparison of the Heun's integration method with the analytical prediction as function of temperature is shown in Fig.~\ref{fig:conv}(c). Here, we consider 1D chain with a length of 500 magnetic moments with nearest-neighbors ferromagnetic exchange and zero anisotropy $\kappa = 0$.

\subsection{Simulation of spinwaves}\label{sec:sim-SW}
To simulate spinwaves we considered a system with a size of $N_1\times N_2 = 500\times500$ magnetic moments. The simulations are carried out in two steps. In the first step, we simulate the dynamics of the system in the external magnetic field $\vec{B}_i= B_0\, \text{sinc}\left(2\pi\omega_0\ \vec{k}\cdot\vec{R}_i\right)\text{sinc}\left[2\pi \left(t-t_0\right)\right]$, where $t_0 = t_\text{sim} / 2$ is a center of temporal part of field profile and $H_0$ is an amplitude of the applied field. The simulations are performed in the low damping regime with $\alpha = 10^{-3}$.

In the second step, we performed a space-time transform for the complex-valued parameter $m_i^\textsc{x}(t)+ i\ m_i^\textsc{y}(t)$. The resulting eigenfrequencies are plotted in Fig.~\ref{fig:disp}.

\begin{figure}[t]
	\includegraphics[width=0.9\columnwidth]{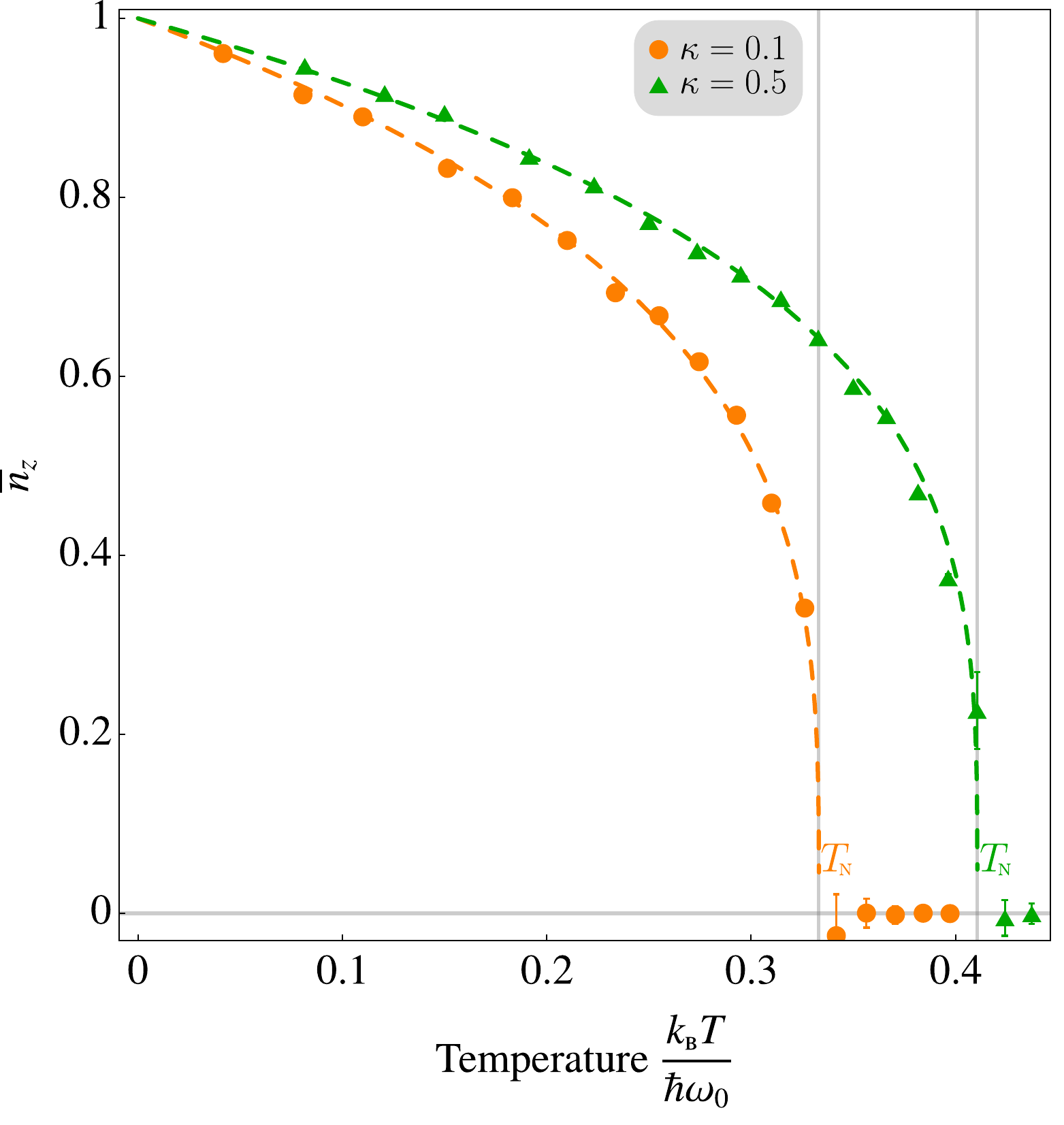}
	\caption{Temperature dependence of perpendicular component of N{\'e}el vector $n_z$ is shown for $\epsilon_x=-0.4$, $\epsilon_y=-0.2$, and for two different anisotropy values. Symbols correspond to the results of numerical simulations. Dashed lines show fitting $\bar{n}_z=\left(1-T/T_\textsc{n}\right)^\alpha$ with (i) $T_\textsc{n}\approx0.33\hbar \omega_0 / k_\textsc{b}$ and $\alpha\approx 0.29$ for $\kappa=0.1$, and (ii) $T_\textsc{n}\approx0.41\hbar \omega_0 / k_\textsc{b}$ and $\alpha=0.27$ for $\kappa=0.5$.}\label{fig:N-vs-T}
\end{figure}

\subsection{Simulation of temperature-induced effects}\label{sec:sim-TS}

Here we consider a system with a size of $N_1\times N_2 = 500\times500$ magnetic moments. In the simulations, we consider the temperature in the range $T\in\left[0.05;\, 0.45\right]\hbar\omega_0/k_\textsc{b}$. The simulations were performed for a low damping regime with $\alpha=10^{-3}$ for a long time scale with $\alpha t\omega_0\gg 1$. The averaged perpendicular net magnetization and N{\'e}el vectors are presented in Figs.~\ref{fig:M-vs-T} and \ref{fig:N-vs-T}, respectively. Note that the averaged in-plane components of the net magnetization vanish.

\section{Fluctuation-induced piezomagnetism and thermal spin conductivity for rutiles}\label{app:rutiles}
Here we consider a modified model of a rutile bilayer film previously proposed in \cite{Gomonay24a}, see Fig~\ref{fig:rutile}.
Magnetic Hamiltonian consists of several contributions: $\mathcal{H}=\mathcal{H}_{\textsc{afm}}+\mathcal{H}_{\textsc{fm}}+\mathcal{H}_{\textsc{alt}}+\mathcal{H}_{\textsc{a}}+\mathcal{H}_{\textsc{z}}$. The leading one is the antiferromagnetic exchange between sublattices 
\begin{equation}
\mathcal{H}_{\textsc{afm}}=J\sum\limits_{\vec{R}_{\vec{n}}}\sum\limits_{\varsigma\in\Xi_{\times}}\vec{m}_1(\vec{R}_{\vec{n}})\cdot\vec{m}_2(\vec{R}_{\vec{n}}+\vec{\delta R}_{\varsigma}),
\end{equation}
where $J>0$ and $\Xi_\times=\{\nearrow,\nwarrow,\swarrow,\searrow\}$ and we introduced the displacement vectors $\vec{\delta R}_{\nearrow}=-\vec{\delta R}_{\swarrow}=\frac{a_0}{2}(\vec{e}_1+\vec{e}_2)$, $\vec{\delta R}_{\nwarrow}=-\vec{\delta R}_{\searrow}=\frac{a_0}{2}(-\vec{e}_1+\vec{e}_2)$. Additionally we take into account the nearest-neighbours ferromagnetic exchange interaction within each of the sublattices
\begin{equation}
\begin{split}
\mathcal{H}_{\textsc{fm}}=&-\frac{J_{\textsc{fm}}}{2}\sum\limits_{\vec{R}_{\vec{n}}}\sum\limits_{\varsigma\in\Xi_{+}}\vec{m}_1(\vec{R}_{\vec{n}})\cdot\vec{m}_1(\vec{R}_{\vec{n}}+\vec{\delta R}_{\varsigma})\\
&-\frac{J_{\textsc{fm}}}{2}\sum\limits_{\vec{R}'_{\vec{n}}}\sum\limits_{\varsigma\in\Xi_{+}}\vec{m}_2(\vec{R}'_{\vec{n}})\cdot\vec{m}_2(\vec{R}'_{\vec{n}}+\vec{\delta R}_{\varsigma}),
\end{split}
\end{equation}
where $J_{\textsc{fm}}>0$ and $\Xi_+=\{\uparrow,\downarrow,\leftarrow,\rightarrow\}$ and the displacement vectors are $\vec{\delta R}_\rightarrow=-\vec{\delta R}_\leftarrow=a_0\vec{e}_x$ and $\vec{\delta R}_\uparrow=-\vec{\delta R}_\downarrow=a_0\vec{e}_y$. The altermagnetic properties are modeled by the next-nearest exchange couplings acting within each of the sublattices
\begin{widetext}
 \begin{equation}
\begin{split}
\mathcal{H}_{\textsc{alt}}=&\frac{\tilde{J}_1}{2}\sum\limits_{\vec{R}_{\vec{n}}}\vec{m}_1(\vec{R}_{\vec{n}})\cdot\bigl[\vec{m}_1(\vec{R}_{\vec{n}}+2\vec{\delta R}_{\searrow})+\vec{m}_1(\vec{R}_{\vec{n}}+2\vec{\delta R}_{\nwarrow})-\vec{m}_1(\vec{R}_{\vec{n}}+2\vec{\delta R}_{\swarrow})-\vec{m}_1(\vec{R}_{\vec{n}}+2\vec{\delta R}_{\nearrow})\bigr]\\
+&\frac{\tilde{J}_2}{2}\sum\limits_{\vec{R}'_{\vec{n}}}\vec{m}_2(\vec{R}'_{\vec{n}})\cdot\bigl[-\vec{m}_2(\vec{R}'_{\vec{n}}+2\vec{\delta R}_{\searrow})-\vec{m}_2(\vec{R}'_{\vec{n}}+2\vec{\delta R}_{\nwarrow})+\vec{m}_2(\vec{R}'_{\vec{n}}+2\vec{\delta R}_{\swarrow})+\vec{m}_2(\vec{R}'_{\vec{n}}+2\vec{\delta R}_{\nearrow})\bigr]
\end{split}
\end{equation}   
\end{widetext}
In contrast to the previously considered model of a rutile film~\cite{Gomonay24a}, here we assume that generally $\tilde{J}_1\ne\tilde{J}_2$. The latter model the action of a stress applied along diagonal directions [110] or [$\bar{1}$10]. We assume that $|\tilde{J}_{1,2}|\ll J+J_{\textsc{fm}}$, thus the ground state is a collinear-compensated one. We also take into account the perpendicular easy-axial anisotropy
\begin{equation}
    \mathcal{H}_{\textsc{a}}=-K\left[\sum\limits_{\vec{R}_{\vec{n}}}m_{1z}^2(\vec{R}_{\vec{n}})+\sum\limits_{\vec{R}'_{\vec{n}}}m_{2z}^2(\vec{R}'_{\vec{n}})\right]
\end{equation}
with $K>0$ and Zeeman interaction with magnetic field $\vec{B}=B\vec{e}_z$
\begin{equation}
    \mathcal{H}_{\textsc{z}}=-\mu_s B\left[\sum\limits_{\vec{R}_{\vec{n}}}m_{1z}(\vec{R}_{\vec{n}})+\sum\limits_{\vec{R}'_{\vec{n}}}m_{2z}(\vec{R}'_{\vec{n}})\right]
\end{equation}

In reciprocal space, Hamiltonian of the spin waves excited over the ground state $\vec{m}_1=\vec{e}_z$, $\vec{m}_2=-\vec{e}_z$ has the structure of \eqref{eq:Hk}
with $\mathcal{A}_{\vec{k}}=\cos\frac{k_xa_0}{2}\cos\frac{k_ya_0}{2}$, $\mathcal{B}_{\vec{k}}=1+\frac{\kappa}{2}+\upsilon(\sin^2\frac{k_xa_0}{2}+\sin^2\frac{k_ya_0}{2})+\delta\epsilon\sin(k_xa_0)\sin(k_ya_0)$, $\mathcal{C}_{\vec{k}}=\bar{\epsilon}\sin(k_xa_0)\sin(k_ya_0)+b$. Here $\kappa$ and $b$ are the same as previously, $\upsilon$, $\bar{\epsilon}$, $\delta\epsilon$ are defined in the main text. The corresponding dispersion relation has the form of \eqref{eq:disp-omega} with $F_{\vec{k}}=\sqrt{\mathcal{B}_{\vec{k}}^2-\mathcal{A}_{\vec{k}}^2}$ and $\Omega_{\vec{k}}^-=\mathcal{C}_{\vec{k}}$. The derivation is analogous to one presented in Appendix~\ref{app:disp}. 

In the low temperature limit, the corresponding asymptotic for $\eta$ coincides with \eqref{eq:eta-f} up to the value of the function $f=-16\bar{\epsilon}(1+\frac{\kappa}{2})/[1+\upsilon(2+\kappa)]^3$. Here we also assumed that $\bar{\epsilon}\sqrt{\kappa}\ll1$. For the parameters of RuO$_2$ listed in supplemental materials of Ref.~\onlinecite{Gomonay24a} we estimate $\eta\approx0.13\mu_{\textsc{b}}/\mathrm{nm}^2$ for the room temperature.

Let us consider the thermal spin conductivity for the rutile model. We limit ourselves to the case $\delta\epsilon=0$, $b=0$, and we also denote $\epsilon=\bar\epsilon$. 
Taking into account the symmetry properties of $\Omega_{\vec{k}}^-$ and $F_{\vec{k}}$ in this case, we obtain from \eqref{eq:sigma-gen} $\sigma_{xx}=\sigma_{yy}=0$ and $\sigma_{xy}=\sigma_{yx}=\sigma$ with
\begin{equation}
    \begin{split}
        &\sigma=-\frac{\gamma k_{\textsc{b}}}{2}\tau_{\text{rlx}}\omega_0\beta'^2\iint\limits_{-\pi}^{\;\;\pi}\frac{\dd q_x\dd q_y}{(2\pi)^2}\\
        &\frac{(F_{\vec{q}}+\Omega_{\vec{q}}^-)(\partial_{q_x}F_{\vec{q}}+\partial_{q_x}\Omega_{\vec{q}}^-)(\partial_{q_y}F_{\vec{q}}+\partial_{q_y}\Omega_{\vec{q}}^-)}{\sinh^2[\frac{\beta'}{2}(F_{\vec{q}}+\Omega_{\vec{q}}^-)]}.
    \end{split}
\end{equation}
Rotation of the reference frame by $\pi/4$ transforms the conductivity tensor $\sigma_{\alpha\beta}$ to \eqref{eq:sigma-tensor}. In the low-temperature limit for the case $\epsilon\ll1$ we estimate
\begin{equation}\label{eq:sigma-ass-rtl}
    \sigma\approx\gamma k_{\textsc{b}}\frac{\omega_0\tau_{\text{rlx}}}{\pi}\Delta^2\epsilon\mathfrak{c}\ln\frac{1}{1-e^{-\beta'\Delta}}.
\end{equation}
Approximation \eqref{eq:sigma-ass-rtl} coincides with \eqref{eq:sigma-ass} up to the  multiplier $\mathfrak{c}=-4/[1+\upsilon(2+\kappa)]$.

\section{Magnetization for low temperature}\label{app:approx}

Based on the identity $1/(e^x-1)=\sum_{n=1}^\infty e^{-nx}$, we present $\sinh(\beta'\Omega_q^-)/\left[\cosh(\beta'F_q)-\sinh(\beta'\Omega_q^-)\right]=1/[e^{\beta'(F_q-\Omega_q^-)}-1]-1/[e^{\beta'(F_q+\Omega_q^-)}-1]=\sum_{n=1}^\infty [e^{-n\beta'(F_q-\Omega_q^-)}-e^{-n\beta'(F_q+\Omega_q^-)}]$. In the low-temperature limit ($\beta'/\Delta\gg1$), the above expression is exponentially localized in the vicinity of $\Gamma$-point. Thus, one can extend the integration over $q_\alpha$ in \eqref{eq:M-den} to infinity. Introducing notations $q_x=q\cos\chi$ and $q_y=q\sin\chi$, we approximate $\Omega_q^-\approx a(\chi)q^2$, $F_q\approx\Delta+b(\chi)q^2$, where $a(\chi)=(\epsilon_x\cos^2\chi-\epsilon_y\sin^2\chi)/8$, and $b(\chi)=[1+(1+\frac{\kappa}{2})(\epsilon_x\cos^2\chi+\epsilon_y\sin^2\chi)]/(8\Delta)$. Now one can easily integrate \eqref{eq:M-den} and obtain $ \mathcal{M}\approx-\frac{g\mu_{\textsc{b}}}{a_0^2}\frac{1}{4\pi^2\beta'}\ln(1-e^{-\beta'\Delta})\int_0^{2\pi}a(\chi)/[b^2(\chi)-a^2(\chi)]\dd\chi$, where we utilize the identity $\sum_{n=1}^\infty e^{-nx}/n=-\ln(1-e^{-x})$. Introducing  $\delta\epsilon=(\epsilon_x-\epsilon_y)/2$ and $\bar{\epsilon}=(\epsilon_x+\epsilon_y)/2$, we writhe the piezomagnetic coefficient in form \eqref{eq:M-asym} with
\begin{subequations}
\begin{align}
\label{eq:eta-f}
 f&=\frac{1}{\pi\delta\epsilon}\int\limits_0^{\pi}\biggl\{\left[1+\left(1+\frac{\kappa}{2}\right)(\bar{\epsilon}+\delta\epsilon\cos\chi)\right]^2\\\nonumber
	&-\Delta^2(\delta\epsilon+\bar{\epsilon}\cos\chi)^2\biggr\}^{-1}(\delta\epsilon+\bar{\epsilon}\cos\chi)\,\dd\chi
 \end{align}
\end{subequations}
In the limit $\delta\epsilon\ll1$, we obtain
\begin{equation}\label{eq:fappr}
	f=\frac{1}{\left\{\left[1+(1+\frac{\kappa}{2})\bar{\epsilon}\right]^2-\Delta^2\bar{\epsilon}^2\right\}^{3/2}}+\mathcal{O}(\delta\epsilon^2).
\end{equation}
For $\bar{\epsilon}\ll1$, one approximates $f\approx1-3(1+\frac{\kappa}{2})\bar{\epsilon}$.

\section{Continuous model and the strain-induced ferrimagnetism}\label{app:cont-model}
In the discrete Hamiltonian \eqref{eq:H}, we perform the Taylor expansion $\vec{m}_\nu(\vec{R}_{\vec{n}}+\delta\vec{R})\approx\vec{m}_\nu(\vec{R}_{\vec{n}})+(\delta\vec{R})_\alpha\partial_\alpha\vec{m}_\nu+\frac12(\delta\vec{R})_\alpha(\delta\vec{R})_\beta\partial_{\alpha\beta}^2\vec{m}_\nu$ and proceed from the summation to integration in the way $\sum_{\vec{R}_{\vec{n}}}(\dots)\to a_0^{-2}\int(\dots)\dd x\dd y$. The continuous approximation of \eqref{eq:H} obtained in this way is $\mathcal{H}=\int\mathscr{H}\dd x\dd y$ with density
\begin{equation}\label{eq:H-cnt}
\begin{split}
    &\mathscr{H}=\frac{1}{a_0^2}\left[4J\,\vec{m}_1\cdot\vec{m}_2+\tilde{J}_y|\vec{m}_1|^2+\tilde{J}_x|\vec{m}_2|^2\right]\\
    &-J\,\partial_\alpha\vec{m}_1\cdot\partial_\alpha\vec{m}_2-\frac{\tilde{J}_y}{2}|\partial_y\vec{m}_1|^2-\frac{\tilde{J}_x}{2}|\partial_x\vec{m}_2|^2\\
    &-\frac{1}{a_0^2}\sum\limits_{\nu=1,2}(Km_{\nu z}^2+B\mu_sm_{\nu z}).
\end{split}    
\end{equation}
In terms of the N{\'e}el $\vec{\mathrm{n}}=\frac12(\vec{m}_1-\vec{m}_2)$ and magnetization  $\vec{\mathrm{m}}=\frac12(\vec{m}_1+\vec{m}_2)$ vectors, we present \eqref{eq:H-cnt} in form
\begin{equation}
    \begin{split}
    \mathscr{H}\approx&\mathscr{H}_{\text{hom}}+\mathcal{A}_{\alpha\beta}\partial_\alpha\vec{\mathrm{n}}\cdot\partial_\beta\vec{\mathrm{n}}+\tilde{\mathcal{A}}_{\alpha\beta}\partial_\alpha\vec{\mathrm{n}}\cdot\partial_\beta\vec{\mathrm{m}}\\
    &-\frac{2}{a_0^2}(K\mathrm{n}_z^2+B\mu_s\mathrm{m}_z).
    \end{split}
\end{equation}
Here $[\mathcal{A}_{\alpha\beta}]=\text{diag}(J-\frac12\tilde{J}_x,J-\frac12\tilde{J}_y)$, and $[\tilde{\mathcal{A}}_{\alpha\beta}]=\text{diag}(\tilde{J}_x,-\tilde{J}_y)$, and we neglected quadratic in $m$ terms except the homogeneous exchange contribution 
\begin{equation}\label{eq:H-hom}
\begin{split}
    \mathscr{H}_{\text{hom}}=\frac{1}{a_0^2}\biggl[&-a_{\textsc{gl}}|\vec{\mathrm{n}}|^2+\frac{b_{\textsc{gl}}}{2}|\vec{\mathrm{n}}|^4\\
    &+4J|\vec{\mathrm{m}}|^2-2(\tilde{J}_x-\tilde{J}_y)\vec{\mathrm{n}}\cdot\vec{\mathrm{m}}\biggr].
\end{split}    
\end{equation}
Here $a_{\textsc{gl}}=4J-\tilde{J}_x-\tilde{J}_y>0$ and additionally we introduce the nonlinear Ginzburg-Landau term with $b_{\textsc{gl}}>0$ which stabilizes the length of the N{\'e}el order parameter. Note that $|\vec{m}_1|\ne|\vec{m}_2|$ if $\tilde{J}_x\ne\tilde{J}_y$, and therefore $\vec{\mathrm{n}}\cdot\vec{\mathrm{m}}\ne0$ in general case. In the leading order in $\tilde{J}_\alpha$ the minimization of $\mathscr{H}_{\text{hom}}$ with respect to $\vec{\mathrm{n}}$ and $\vec{\mathrm{m}}$ results in
\begin{equation}
    \vec{\mathrm{m}}=\frac{\epsilon_x-\epsilon_y}{4}\vec{\mathrm{n}},\qquad|\vec{\mathrm{n}}|\approx\sqrt{\frac{a_{\textsc{gl}}}{b_{\textsc{gl}}}},
\end{equation}
where $\epsilon_\alpha=\tilde{J}_\alpha/J$. The corresponding magnetic moment density is $\mathcal{M}=\eta_{\textsc{fm}}(\epsilon_x-\epsilon_y)$ where $\eta_{\textsc{fm}}=\mu_s|\vec{\mathrm{n}}|/(2a_0^2)$ is the piezomagnetic coupling constant which originates from the strain-induced ferrimagnetism. In contrast to $\eta_{\textsc{fm}}$, the fluctuations related piezomagnetic constant $\eta$ does not depend on $\mu_s$, see Fig.~\ref{fig:M-vs-T}. This is a consequence of the fact that magnetic moment of a magnon $\pm\gamma\hbar$ does not depend on $\mu_s$. As a result, for materials with $\mu_s\ll\mu_{\textsc{b}}$, we expect that $\eta\gg\eta_{\textsc{fm}}$ in the limit of high temperatures $T\lesssim T_{\textsc{n}}$. In this temperature regime, $\eta$ has the highest value while $|\vec{\mathrm{n}}|$ decreases leading also to the additional decrease of $\eta_{\textsc{fm}}$.

\section{A simple 3D generalization of the checkerboard model}\label{app:3D}
As a three-dimensional generalization of the checkerboard model we consider a stacked system composed of the 2D checkerboard layers shown in Fig.~\ref{fig:checkerboard} with ferromagnetic inter-layer exchange coupling. The 3D Hamiltonian is
\begin{equation}
\begin{split}
    \mathcal{H}^{\text{3D}}\!=\!N\mathcal{H}-&J_z\biggl[\sum\limits_{\vec{R}_{\vec{n}}^{\text{3D}}}\vec{m}_1(\vec{R}_{\vec{n}}^{\text{3D}})\!\cdot\!\vec{m}_1(\vec{R}_{\vec{n}}^{\text{3D}}+c_0\vec{e}_z)\\
    +&\sum\limits_{\vec{R}'^{\text{3D}}_{\vec{n}}}\vec{m}_2(\vec{R}'^{\text{3D}}_{\vec{n}})\!\cdot\!\vec{m}_2(\vec{R}'^{\text{3D}}_{\vec{n}}+c_0\vec{e}_z)\biggr],
    \end{split}
\end{equation}
where $J_z>0$, and $c_0$ is the inter-layer distance. Vectors
 $\vec{R}^{\text{3D}}_{\vec{n}}=a_0(n_x\vec{e}_x+n_y\vec{e}_y)+c_0n_z\vec{e}_z$ and $\vec{R}'^{\text{3D}}_{\vec{n}}=\vec{R}^{\text{3D}}_{\vec{n}}-\frac{a_0}{2}(\vec{e}_x+\vec{e}_y)$ determine the positions of magnetic moments belonging to the first and second sublattices, respectively. Hamiltonian $\mathcal{H}$ is determined in \eqref{eq:H}, and $N$ is the number of layers. 
 
 In reciprocal space, the Hamiltonian of the spin waves excited over the ground state $\vec{m}_1=\vec{e}_z$, $\vec{m}_2=-\vec{e}_z$ coincides with \eqref{eq:Hk} up to replacement $\mathcal{B}_{\vec{k}}\to\mathcal{B}_{\vec{k}}+\upsilon_z\sin^2\frac{k_zc_0}{2}$ with $\upsilon_z=J_z/J$. The corresponding dispersion relation has the form of Eq.~\eqref{eq:disp-omega} with $F_{\vec{k}}=\sqrt{(1+\frac{\kappa}{2}-\Omega_{\vec{k}}^++\upsilon_z\sin^2\frac{k_zc_0}{2})^2-\cos^2\frac{k_xa_0}{2}\cos^2\frac{k_ya_0}{2}}$ and $\Omega_{\vec{k}}^\pm$ is defined in \eqref{eq:disp-Omega}.

 Similarly to \eqref{eq:M-den}, we obtain the 3D generalization of the magnetic moment density
 \begin{equation}\label{eq:M-den3D}
	\mathcal{M}^{\text{3D}}=\frac{g\mu_{\textsc{b}}}{a_0^2c_0}\!\!\iiint\limits_{-\pi}^{\;\;\;\;\;\pi}\!\!\frac{\dd \vec{q}}{(2\pi)^3}\frac{\sinh\left(\beta'\Omega_{\vec{q}}^-\right)}{\cosh\left(\beta'F_{\vec{q}}\right)-\cosh\left(\beta'\Omega_{\vec{q}}^-\right)},
\end{equation}
where $\dd\vec{q}=\dd q_x\dd q_y\dd q_z$ and $q_z=k_zc_0$. Since $|\epsilon_\alpha|\ll1$, in the leading order in $\epsilon_\alpha$ we approximate
 \begin{equation}\label{eq:M-den3D-approx}
	\mathcal{M}^{\text{3D}}\approx\frac{g\mu_{\textsc{b}}}{a_0^2c_0}\frac{\beta'}{8}\!\!\iiint\limits_{-\pi}^{\;\;\;\;\;\pi}\!\!\frac{\dd \vec{q}}{(2\pi)^3}\frac{\epsilon_xq_x^2-\epsilon_yq_y^2}{\cosh\left(\beta'F_{\vec{q}}^0\right)-1},
\end{equation}
where $F_{\vec{k}}^0=\sqrt{(1+\frac{\kappa}{2}+\upsilon_z\sin^2\frac{k_zc_0}{2})^2-\cos^2\frac{k_xa_0}{2}\cos^2\frac{k_ya_0}{2}}$.
In the low-temperature limit, meaning $k_{\textsc{b}}T\ll(\hbar\omega_0)^2/(\hbar\omega_{\text{gap}})$ and $k_{\textsc{b}}T\ll\upsilon_z(\hbar\omega_0)^2/(\hbar\omega_{\text{gap}})$ with $\omega_{\text{gap}}=\omega_0\Delta$, from \eqref{eq:M-den3D-approx} we obtain the following estimation for the piezomagnetic coefficient $\eta^{\text{3D}}=\mathcal{M}^{\text{3D}}/(\epsilon_x-\epsilon_y)$:
\begin{equation}\label{eq:eta-3D}
    \eta^{\text{3D}}\approx\frac{g\mu_{\textsc{b}}}{a_0^2c_0}\left(\frac{8}{\pi}\right)^{\frac{3}{2}}\frac{8\Delta^{5/2}}{\sqrt{\upsilon_z(2+\kappa)}}\frac{1}{\beta'^{3/2}}\text{Li}_{\frac{3}{2}}e^{-\beta'\Delta},
\end{equation}
where $\text{Li}_s(x)=\sum_{n=1}^{\infty}x^n/n^{s}$ is the polylogarithm function.

In a gapless case ($\Delta=0$), in the low-temperature limit $k_{\textsc{b}}T\ll\sqrt{\upsilon_z}\hbar\omega_0$, we obtain  estimation \eqref{eq:eta-3D-gpls}
from \eqref{eq:M-den3D-approx}.

Let us consider the spin conductivity tensor. We start from three-dimensional generalization of \eqref{eq:sigma-gen}:
\begin{equation}\label{eq:sigma-gen3D}
    \begin{split}
        &\sigma_{ij}^{\mathrm{3D}}=-\frac{\gamma k_{\textsc{b}}}{4}\frac{\tau_{\text{rlx}}\omega_0}{L_xL_yL_z}\beta'^2\sum\limits_{\vec{k}\in1.\text{BZ}}\sum\limits_{\nu=\pm1}\\
        &\nu\frac{(F_{\vec{k}}+\nu\Omega_{\vec{k}}^-)(\partial_{k_i}F_{\vec{k}}+\nu\partial_{k_i}\Omega_{\vec{k}}^-)(\partial_{k_j}F_{\vec{k}}+\nu\partial_{k_j}\Omega_{\vec{k}}^-)}{\sinh^2[\frac{\beta'}{2}(F_{\vec{k}}+\nu\Omega_{\vec{k}}^-)]},
    \end{split}
\end{equation}
where $i,j\in\{x,y,z\}$ and the summation is performed over 3D Brillouin zone. We consider the case $\epsilon_x=\epsilon_y=\epsilon$. Taking into account the symmetries of $F_{\vec{k}}$ and $\Omega_{\vec{k}}^-$ discussed in Appendix~\ref{app:sigma} and also that $\partial_{k_z}F_{\vec{k}}$ is an odd function of $k_z$ and $\partial_{k_z}\Omega^-_{\vec{k}}=0$, we obtain that all components of $\sigma_{ij}^{\mathrm{3D}}$ vanish except $-\sigma_{xx}^{\mathrm{3D}}=\sigma_{yy}^{\mathrm{3D}}=\sigma^{\mathrm{3D}}$ with
\begin{equation}\label{eq:sigma3D}
\begin{split}
    &\sigma^{\mathrm{3D}}=\frac{\gamma k_{\textsc{b}}}{4}\tau_{\text{rlx}}\omega_0\frac{\beta'^2}{c_0}\int\frac{\dd\vec{q}}{(2\pi)^3}\\
    &\frac{(F_{\vec{q}}+\Omega^-_{\vec{q}})\left[\left(\mathfrak{f}_y+\frac{\epsilon}{4}\right)^2\sin^2q_x-\left(\mathfrak{f}_x-\frac{\epsilon}{4}\right)^2\sin^2q_y\right]}{\sinh^2\left[\frac{\beta'}{2}(F_{\vec{q}}+\Omega^-_{\vec{q}})\right]}.
    \end{split}
\end{equation}
Here $\mathfrak{f}_\alpha=\left[\cos^2\frac{q_\alpha}{2}-\epsilon(1+\frac{\kappa}{2}-\Omega^+_{\vec{q}}+\upsilon_z\sin^2\frac{q_z}{2})\right]/(4F_{\vec{q}})$. In the low temperature limit $k_{\textsc{b}}T\ll\sqrt{\upsilon_z}\hbar\omega_0$, one approximates \eqref{eq:sigma3D} as \eqref{eq:sigma-3D}.

\section{Tensor of thermal spin conductivity for checkerboard model}\label{app:sigma}

In the general case, the diagonal terms of tensor \eqref{eq:sigma-gen} for the checkerboard model are as
follows
\begin{equation}
\begin{split}
    &\sigma_{xx}=-\gamma k_{\textsc{b}}\beta'^2\frac{\tau_{\text{rlx}}\omega_0}{64}\iint\limits_{-\pi}^\pi\frac{\dd q_x\dd q_y}{(2\pi)^2}\sin^2q_x\\
    &\times\Biggl\{\frac{(F_q+\Omega_q^-)\left[\frac{1}{F_q}\left(\cos^2\frac{q_y}{2}-\epsilon_x\left(1+\frac{\kappa}{2}-\Omega_q^+\right)\right)+\epsilon_x\right]^2}{\sinh^2\left[\frac{\beta'}{2}(F_q+\Omega_q^-)\right]}\\
    &-\frac{(F_q-\Omega_q^-)\left[\frac{1}{F_q}\left(\cos^2\frac{q_y}{2}-\epsilon_x\left(1+\frac{\kappa}{2}-\Omega_q^+\right)\right)-\epsilon_x\right]^2}{\sinh^2\left[\frac{\beta'}{2}(F_q-\Omega_q^-)\right]}
    \Biggr\}
    \end{split}
\end{equation}
and 
\begin{equation}
\begin{split}
    &\sigma_{yy}=-\gamma k_{\textsc{b}}\beta'^2\frac{\tau_{\text{rlx}}\omega_0}{64}\iint\limits_{-\pi}^\pi\frac{\dd q_x\dd q_y}{(2\pi)^2}\sin^2q_y\\
    &\times\Biggl\{\frac{(F_q+\Omega_q^-)\left[\frac{1}{F_q}\left(\cos^2\frac{q_x}{2}-\epsilon_y\left(1+\frac{\kappa}{2}-\Omega_q^+\right)\right)-\epsilon_y\right]^2}{\sinh^2\left[\frac{\beta'}{2}(F_q+\Omega_q^-)\right]}\\
    &-\frac{(F_q-\Omega_q^-)\left[\frac{1}{F_q}\left(\cos^2\frac{q_x}{2}-\epsilon_y\left(1+\frac{\kappa}{2}-\Omega_q^+\right)\right)+\epsilon_y\right]^2}{\sinh^2\left[\frac{\beta'}{2}(F_q-\Omega_q^-)\right]}
    \Biggr\}.
    \end{split}
\end{equation}


\end{document}